\documentclass[12pt]{article}
\usepackage{amsfonts}
\usepackage{amssymb}
\usepackage{graphicx}
\usepackage{amsmath}
\usepackage{makeidx}
\usepackage{indentfirst}
\usepackage[T1]{fontenc}
\usepackage[utf8]{inputenc}

\newcounter{resultnum}[section]
\newcounter{conclusionnum}[section]
\newcounter{conditionnum}[section]
\newcounter{conjecturenum}[section]
\newcounter{examplenum}[section]
\newcounter{exercisenum}[section]
\newcounter{lemmanum}[section]
\newcounter{notationnum}[section]
\newcounter{theoremnum}[section]
\newcounter{definitionnum}[section]
\newcounter{corollarynum}[section]
\newcounter{remarknum}[section]
\newcounter{propositionnum}[section]
\newcounter{acknowledgementnum}[section]
\newcounter{algorithmnum}[section]
\newcounter{axiomnum}[section]
\newcounter{casenum}[section]
\newcounter{claimnum}[section]
\newcounter{summarynum}[section]
\newcounter{problemnum}[section]
\begin{document}

\title{Space-Time Quasicrystal Structures and Inflationary and Late Time
Evolution Dynamics in Accelerating Cosmology}
\date{October 2, 2018}
\author{ \textbf{\Large Sergiu I. Vacaru}\thanks{\textit{The address for
post correspondence:\ 140 Morehampton Rd, Donnybrook, Dublin 04, Ireland,
D04 N2C0;} \newline
emails: sergiu.vacaru@gmail.com ; sergiuvacaru@mail.fresnostate.edu } \\
{\qquad } \\
{\textit{Physics Department, California State University at Fresno, CA
93740, USA}} \\
{and } {\ \textit{\ Project IDEI, University "Al. I. Cuza" Ia\c si, Romania}}%
{\qquad } }
\maketitle

\begin{abstract}
We construct new classes of cosmological solutions in modified and Einstein
gravity theories encoding space-time quasicrystal, STQC, configurations
modeled by nonlinear self-organized and pattern forming quasi-periodic
structures. Such solutions are defined by generic off-diagonal locally
anisotropic and inhomogeneous metrics depending via generating and
integration functions on all spacetime coordinates. There are defined
nonholonomic variables and conditions for the generating/integration
functions and sources for effective descriptions, or approximations, as
"quasi" Friedmann-Lama\^{\i}tre-Robertson-Walker (FLRW) metrics. Such (off-)
diagonal STQC-FLRW configurations contain memory on nonlinear classical
and/or quantum interactions and may describe new acceleration cosmology
scenarios. For special time-periodic conditions on nonlinear gravitational
and matter field interactions, we can model at cosmological scales certain
analogous of time crystal like structures originally postulated by Frank
Wilczek in condensed matter physics. We speculate how STQC quasi-FLRW
configurations could explain modern cosmology data and provide viable
descriptions for the inflation and structure formation in our Universe.
Finally, it is discussed systematically and critically how a unified
description of inflation with dark energy era can be explained by (modified)
cosmological STQC-scenarios.

\vskip3pt

\textbf{Keywords:}\ geometric and condensed matter methods in modern
cosmology; modified gravity theories; space-time quasiperiodic, aperiodic
and quasicrystal cosmological structures; post modern inflation paradigm;
accelerating cosmology; dark energy and dark matter.

\vskip3pt

MSC:\ 83C45, 83C99, 53D55, 53B40, 53B35

PACS:\ 47.54.-r; 05.45.-a; 47.35.Pq; 47.52.+j; 61.50.Ah; 61.44.Br; 05.20.-y;
64.70.D 
\end{abstract}

\tableofcontents



\section{Introduction}

This work provides a concise introduction into the cosmology with space-time
quasi-periodic and pattern forming structures. For the latest developments
on quasiperiodc cosmological models, we cite \cite%
{cosmv2,eegrg,cosmv3,rajpvcosm,amaral17,vbubuianu17} and references therein.
The approach involves cosmological scenarios with locally anisotropic and
inhomogenous gravitational and matter field interactions in general
relativity, GR, and modified gravity theories MGTs. Such results on $f(R)$
cosmology and various generalizations are reviewed in \cite%
{capoz,nojod1,gorbunov,clifton,basilakos,linde2,gwyn,sami,mavromatos,nojiri,vagnozzi}%
. Hence we focus on cosmological models which can be described effectively
by "quasi" Friedmann-Lama\^{\i}tre-Roberston-Walker, FLRW, metrics encoding
space-time quasicrystal structures, STQC, modelled at cosmological scales.
There will be written in brief and used such terms: quasicrystal, QC; time
crystal, TC; time quasicrystal,\ TQC; time crystal, TC, and similarly for
corresponding nonlinear structures and space-time configurations described
by solutions of classical or quantum equations not only in condensed matter
physics but also in gravity theories, astrophysics and cosmology.

We show that nonlinear gravitational and matter field interactions define
STQC structures which are quasiperiodic cosmological analogous to TC and QC.
In our approach, we generalize the concepts of classical and quantum TCs
developed in A. Shapere and F. Wilczek works \cite%
{shapere12,wilczek12,wilczek13a,wilczek13b}, see references therein on other
concepts of TC elaborated for biological systems, quantum coherence, and
cosmology, and \cite{shapere17,sacha18} for a review and recent results.
Here we note that two research groups were able already to create
independently TCs in condensed matter physics \cite%
{li12,yao17,choi17,zhang17}.

Let us explain in brief the difference and similarity between TC and QC
structures in cosmology and condensed matter physics. Originally, A. Shapere
and F. Wilczek \cite{shapere12} elaborated on classical TCs with properties
which are time analogues of crystalline spatial order. They studied
dynamical systems in the lowest energy state when within orbits of broken
symmetry there are modelled interactions with traveling density waves. In a
partner work \cite{wilczek12}, F. Wilczek proposed the existence of a new
state of matter called "quantum time crystals", QTC, for quantum mechanical
systems defined with a ground state displaying a time-dependent behaviour
(periodic oscillation) of some physical observables. If normal crystals
exhibit broken translation symmetry, the concept of TC was elaborated
theoretically for classical and quantum many-body systems with
self-organization in time. Such systems describe a periodic motion with
possible jumps or periodic oscillations after a spontaneously breaking of
time translation symmetry. The works on TCs and QTCs stimulated both a
considerable interest and controversial debates. Several papers claiming
that space-time crystals are impossible were published \cite%
{bruno13a,bruno13b}. In a series of subsequent works \cite%
{nozieres13,volovik13}, certain models of time translation symmetry breaking
were elaborated and studied in details. That led to a conclusion that
quantum TCs in equilibrium thermodynamical states are not possible \cite%
{watanabe15}. As a consequence, several models of TCs which avoid the
equilibrium no-go arguments were elaborated \cite{wilczek13b,yoshii15}, for
instance, for time translation symmetry breaking of non-equilibrium systems
\cite{else16,khemani16,yao17}.

The first goal of this paper is to prove that STQC, TQC and TC structures
can be generated as (off-) diagonal locally anisotropic and inhomogeneous
cosmological solutions in MGTs and GR. In particular, such (quasi) periodic
gravitational and matter field structures are generated by a spontaneous
broken gravitational vacuum metrics $g_{\alpha \beta }(...,t)$ with
dependencies on a time like coordinate $t.$ We shall analyze subclasses of
generating functions for off-diagonal metrics, or certain quasi FLRW
configurations depending only on $t,$ or modeling QC structures as in \cite%
{amaral17,vbubuianu17}. For corresponding functional and parametric
dependencies, such gravitational and matter field solutions encode data on
STQC. The second our goal is to apply our geometric methods and new classes
of cosmological solutions for elaborating new scenarios of inflation with
(or induced by) STQC structure and analyse possible implications in dark
energy and dark matter physics.

Our work is organized as follow. In section \ref{s2}, we formulate the
gravitational and matter field equations for STQCs in MGTs and GR. There are
provided necessary parameterizations of cosmological metrics, coefficient
formulas, geometric calculus and approximations. Section \ref{s3} is devoted
to a geometric method of constructing exact cosmological solutions for STQC
configurations in general locally anisotropic forms and/or with off-diagonal
time dependence or quasi FLRW limits. The inflationary STQC dynamics is
studied in section \ref{s4}. Then, we consider models of late STQC dynamics
and dark energy in section \ref{s5}. Finally, conclusions are formulated in
section \ref{s6}.

\section{Time quasi-periodic configurations in (modified) gravity}

\label{s2}We outline certain advanced geometric methods \cite%
{cosmv2,eegrg,cosmv3,rajpvcosm,amaral17,vbubuianu17} which can be applied
for modeling space-time configurations with STQC structure. Then, there are
formulated basic equations defining space-time QCs and the gravitational and
matter field equations for STQCs.

\subsection{2+2 and 3+1 splitting and distorted (non) linear connections}

Let us consider a four dimensional, 4d, spacetime manifold $\mathbf{V}$
modelled as a Lorentz manifold enabled with a metric $\mathbf{g}$ of
pseudo--Riemannian signature $(+++-).$ For a conventional 2+2 splitting, the
local coordinates are denoted $u^{\gamma }=(x^{k},y^{c}),$ or $u=(x,y),$
with indices $i,j,k,...=1,2$ and $a,b,c,...=3,4.$ We can consider a 3+1
splitting when $u^{4}=y^{4}=t$ is a time like coordinate and $u^{\grave{%
\imath}}=(x^{i},y^{3}),$ where spacelike coordinates run values $\grave{%
\imath},\grave{j},\grave{k},...=1,2,3.$ On respective tangent and cotangent
Lorentz bundles, $T\mathbf{V}$ and (dual) $T^{\ast }\mathbf{V,}$ the local
bases/ frames are written as frame (equivalently, tetrad, or vierbein)
transforms\ related to local coordinate bases, $e_{\alpha }=e_{\alpha \
}^{~\alpha ^{\prime }}(u)\partial _{\alpha ^{\prime }},$ for $\partial
_{\alpha ^{\prime }}=\partial /\partial u^{\alpha ^{\prime }},$ and cobases $%
e^{\alpha }=e_{\ \alpha ^{\prime }}^{\alpha }(u)du^{\alpha ^{\prime }}.$

A nonholonomic 2+2 splitting is determined by a nonlinear connection,
N-connection, structure $\mathbf{N}$ defined as a conventional horizontal,
h, and vertical, v, decomposition. It is stated by a respective nonholonomic
(equivalently, anholonomic, i.e. nonintegrable) distribution,
\begin{equation}
\mathbf{N:\quad }T\mathbf{V}=hT\mathbf{V}\oplus vT\mathbf{V}  \label{ncon}
\end{equation}%
into respective h- and v-subspaces, $hT\mathbf{V}$ and $vT\mathbf{V.}$%
\footnote{%
Such formulas can be written in local form, $\mathbf{N}=N_{i}^{a}\frac{%
\partial }{\partial y^{a}}\otimes dx^{i},$ using the N-connection
coefficients $N_{i}^{a};$ we shall use the Einstein convention on summation
on up-low repeating indices if the contrary will not be stated; the
geometric objects on $\mathbf{V}$ will be labeled by "bold face" symbols if
they can be written in N-adapted form.} Any N-connection structure defines
respective classes of N--adapted (co) frames, $\mathbf{e}_{\alpha }=(\mathbf{%
e}_{i},e_{a}),$\footnote{\label{ftnnonhcoef}a local basis $\mathbf{e}%
_{\alpha }$ is nonholonomic if the commutators $\mathbf{e}_{[\alpha }\mathbf{%
e}_{\beta ]}:= \mathbf{e}_{\alpha}\mathbf{e}_{\beta }-\mathbf{e}_{\beta }%
\mathbf{e}_{\alpha }=C_{\alpha \beta }^{\gamma }(u)\mathbf{e}_{\gamma }$
contain nontrivial anholonomy coefficients $C_{\alpha \beta
}^{\gamma}=\{C_{ia}^{b}=\partial_{a}N_{i}^{b},C_{ji}^{a}= \mathbf{e}%
_{j}N_{i}^{a}-\mathbf{e}_{i}N_{j}^{a}\}$} and dual frames, $\mathbf{e}%
^{\alpha }=(x^{i},\mathbf{e}^{a}),$
\begin{eqnarray}
\mathbf{e}_{i} &=&\partial /\partial x^{i}-N_{i}^{a}(u)\partial /\partial
y^{a},e_{a}=\partial _{a}=\partial /\partial y^{a},  \label{nder} \\
e^{i} &=&dx^{i},\mathbf{e}^{a}=dy^{a}+N_{i}^{a}(u^{\gamma })dx^{i}.  \notag
\end{eqnarray}

Any (pseudo) Riemannian metric on $\mathbf{V}$ can be parameterized as a
distinguished metric, d--metric,
\begin{eqnarray}
\mathbf{g} &=&\mathbf{g}_{\alpha \beta }(u)\mathbf{e}^{\alpha }\otimes
\mathbf{e}^{\beta }=g_{i}(x^{k})dx^{i}\otimes dx^{i}+h_{a}(x^{k},y^{b})%
\mathbf{e}^{a}\otimes \mathbf{e}^{b}  \label{odans} \\
&=&\mathbf{g}_{\alpha ^{\prime }\beta ^{\prime }}(u)\mathbf{e}^{\alpha
^{\prime }}\otimes \mathbf{e}^{\beta ^{\prime }},\mbox{ for }\mathbf{g}%
_{\alpha ^{\prime }\beta ^{\prime }}(u)=\mathbf{g}_{\alpha \beta }\mathbf{e}%
_{\ \alpha ^{\prime }}^{\alpha }\mathbf{e}_{\ \beta ^{\prime }}^{\beta }.
\notag
\end{eqnarray}

A distinguished connection, d--connection, (equivalently, a d-covariant
derivative) $\mathbf{D}=\{\mathbf{D}_{\alpha
}\}=(hD=\{D_{i}\},vD=\{D_{a}\}), $ is defined as a linear connection,
preserving the N--connection splitting $\mathbf{N}$ (\ref{ncon}) under
parallel transports. The d-connection coefficients $\mathbf{D}=\{\mathbf{%
\Gamma }_{\ \beta \gamma }^{\alpha }=(L_{\ jk}^{i},\acute{L}_{\ bk}^{a};
\acute{C}_{\ jc}^{i},C_{\ bc}^{a})\},$ with respective h- and v-covariant
derivatives, $\ _{h}\mathbf{D}=\{(L_{\ jk}^{i},\acute{L}_{\ bk}^{a})\}$ and $%
\ _{v}\mathbf{D}=\{(\acute{C}_{\ jc}^{i},C_{\ bc}^{a})\},$ are computed in
N-adapted form with respect to frames (\ref{nder}).

A nonholonomic metric-affine Lorentz manifold is defined by data $\left(
\mathbf{V,N,D,g}\right) ,$ where $\mathbf{g}$ is a Lorentz metric. In
general, a d-connection $\mathbf{D}=\nabla +\mathbf{Z,}$ with a distortion
tensor $\mathbf{Z,}$ is different from the standard Levi-Civita connection,
LC-connection, $\nabla ,$ which (by definition) is metric compatible and
with zero torsion.\footnote{%
Using standard methods of differential geometry, we can define and compute
the N-adapted coefficients of the d-tensors for curvature $\mathcal{R}=\{%
\mathbf{R}_{\ \beta \gamma \delta }^{\alpha }\},$ torsion $\mathcal{T}= \{%
\mathbf{T}_{\beta \gamma }^{\alpha }\},$ and nonmetricity $\mathcal{Q}=\{%
\mathbf{Q}_{\alpha \beta \gamma }:=\mathbf{D}_{\alpha }\mathbf{g}_{\beta
\gamma }\}.$}

On a nonholonomic Lorentz manifold $\left( \mathbf{V,N}\right) ,$ we can
work equivalently with two important linear connections:
\begin{equation}
(\mathbf{g,N})\rightarrow \left\{
\begin{array}{cc}
\mathbf{\nabla :} & \mathbf{\nabla g}=0;\ ^{\nabla }\mathcal{T}=0,%
\mbox{\
for  the Levi-Civita,  LC, -connection } \\
\widehat{\mathbf{D}}: & \widehat{\mathbf{D}}\mathbf{g}=0;\ h\widehat{%
\mathcal{T}}=0,v\widehat{\mathcal{T}}=0,hv\widehat{\mathcal{T}}\neq 0,%
\mbox{
for the canonical d--connection  }.%
\end{array}%
\right.  \label{twocon}
\end{equation}%
We note that both $\mathbf{\nabla }$ and $\widehat{\mathbf{D}}$ are defined
by the same metric structure $\mathbf{g}$ which allows to model any (pseudo)
Riemannian geometry as an effective a (pseudo) Riemann--Cartan geometry with
nonholonomically induced torsion structure $\widehat{\mathcal{T}}=\{\widehat{%
\mathbf{T}}_{\beta \gamma }^{\alpha }[N_{i}^{a},\mathbf{g}_{\beta \gamma}]\}$%
. The coefficients $\widehat{\mathbf{T}}_{\beta \gamma }^{\alpha }$ are
determined by nontrivial anholonomy coefficients $C_{\alpha \beta }^{\gamma
} $ and the coefficients of the metric and N-connection. It is always
possible to imposing additional nonholonomic constraints, or limits, when, $%
\widehat{\mathcal{T}}=0$ and $\widehat{\mathbf{D}}_{\mid \widehat{\mathcal{T}%
}=0}= \mathbf{\nabla .}$\footnote{%
The curvature tensors of both linear connections are computed in standard
forms, respectively, for $\widehat{\mathbf{D}}$ and $\nabla ,$ when $%
\widehat{\mathcal{R}}=\{\widehat{\mathbf{R}}_{\beta \gamma \delta }^{\alpha
}\}$ and $\ ^{\nabla }\mathcal{R}=\{R_{\ \beta \gamma \delta}^{\alpha }\}$.
The corresponding Ricci tensors are $\ \widehat{\mathcal{R}}ic=\{\widehat{%
\mathbf{R}}_{\ \beta \gamma }:=\widehat{\mathbf{R}}_{\ \alpha \beta \gamma
}^{\gamma }\}$ and $Ric=\{R_{\ \beta \gamma }:=R_{\ \alpha \beta \gamma
}^{\gamma }\},$ where $\widehat{\mathcal{R}}ic$ is characterized by $h$-$v$
N-adapted coefficients, $\widehat{\mathbf{R}}_{\alpha \beta }=\{\widehat{R}%
_{ij}:=\widehat{R}_{\ ijk}^{k},\ \widehat{R}_{ia}:=-\widehat{R}_{\
ika}^{k},\ \widehat{R}_{ai}:=\widehat{R}_{\ aib}^{b},\ \widehat{R}_{ab}:=%
\widehat{R}_{\ abc}^{c}\}.$ There are also two different scalar curvatures, $%
\ R:=\mathbf{g}^{\alpha \beta }R_{\alpha \beta }$ and $\ \widehat{\mathbf{R}}%
:=\mathbf{g}^{\alpha \beta }\widehat{\mathbf{R}}_{\alpha \beta }=g^{ij}%
\widehat{R}_{ij}+g^{ab}\widehat{R}_{ab}$.}

The canonical d-connection $\widehat{\mathbf{D}}$ is very important because
it allows a general decoupling and integration of modified Einstein
equations. In our approach we consider systems of nonlinear partial
differential equations, PDEs, in general forms and do not reduce them to
some particular cases of systems of nonlinear ordinary differential
equations, ODEs. We work with various classes of off-diagonal metrics $%
\mathbf{g}_{\beta \gamma }(x^{k},y^{c})$ depending, in principle, on all
spacetime coordinates, see details in \cite{vbubuianu17} and references
therein. Having found a class of solutions, for instance, for locally
anisotropic cosmological metrics $\mathbf{g}_{\beta \gamma }(x^{k},t))$
determined by respective sets of generating and integration functions, we
can impose additional constraints which allow us to extract
LC-configurations with $\widehat{\mathbf{D}}_{\mid \widehat{\mathcal{T}}=0}=%
\mathbf{\nabla }$ and (as particlar cases) cosmological solutions with $%
\mathbf{g}_{\beta \gamma }(t)$. Generalized (off-)diagonal classes of
solutions (for instance, with STQC structure) can not be constructed if we
work from the beginning with $\mathbf{\nabla }$ for some high symmetric
ansatz for metrics - this is a general property of the systems of nonlinear
PDEs.

\subsection{Space-time QC structures}

In this subsection, we provide three examples how time quasiperiodic
structures can be defined in a curved space-time. We consider one and two
dimensional TQCs with time crystal equations generalizing those introduced
in \cite{shapere17}); three dimensional QC structures studied in \cite%
{vbubuianu17,amaral17}; and mixed configurations resulting in STQC. We
emphasize that in this work we use a different system of notation for
partial derivatives when, for instance, $\partial q/\partial
x^{i}=\partial_{i}q,$ $\partial q/\partial y^{3}=\partial _{3}q,$ and $%
\partial q/\partial y^{4}=\partial _{4}q=\partial _{4}q=q^{\bullet },$ for a
function $q(x^{i},y^{3},t).$ A prime will be used for any functional
derivative like $F^{\prime }(R)=dF/dR,$ for instance, in a modified gravity
with $F(R),$ where is a Ricci scalar.

\subsubsection{One dimensional relativistic time quasicrystal structures}

Such a structure on a space-time $(\mathbf{V,g,N)}$ can be modelled by a
scalar field $\varsigma (x^{i},y^{a})$ and a Lagrange density
\begin{equation}
\acute{L}[\varsigma (x^{i},y^{a})]=\frac{1}{48}(\mathbf{g}^{\alpha \beta }(%
\mathbf{e}_{\alpha }\varsigma )(\mathbf{e}_{\beta }\varsigma ))^{2}-\frac{1}{%
4}\mathbf{g}^{\alpha \beta }(\mathbf{e}_{\alpha }\varsigma )(\mathbf{e}%
_{\beta }\varsigma )-\acute{V}(\varsigma )  \label{1tqclg}
\end{equation}%
where $\acute{V}(\varsigma )$ is a nonlinear potential and $\mathbf{e}%
_{\alpha }$ (\ref{ncon}) are N-adapted partial derivatives. This $\acute{L}$
results in variational motion equations of type
\begin{equation}
\lbrack \frac{1}{2}\mathbf{g}^{\alpha \beta }(\mathbf{e}_{\alpha }\varsigma
)(\mathbf{e}_{\beta }\varsigma )-1](\widehat{\mathbf{D}}^{\gamma }\widehat{%
\mathbf{D}}_{\gamma }\varsigma )=2\frac{\partial \acute{V}}{\partial
\varsigma }.  \label{1tqc}
\end{equation}%
We say that a field $\varsigma $ defines a one dimensional (1-d) time
quasicrystal structure, 1-TQC, if it is a solution of the equation (\ref%
{1tqc}).

For non-relativistic approximations with $g_{\alpha \beta }=[1,1,1,-1]$ and $%
\varsigma \rightarrow \varsigma (t),$ we obtain an effective Lagrange
density
\begin{equation}
\acute{L}[\varsigma (x^{i},y^{a})]\rightarrow \acute{L}[\varsigma (t)]=\frac{%
1}{12}(\varsigma ^{\bullet })^{4}-\frac{1}{2}(\varsigma ^{\bullet })^{2}-%
\acute{V}(\varsigma ),  \label{1tqclh}
\end{equation}%
leading to an effective energy $E=\frac{1}{4}[(\varsigma ^{\bullet
})^{2}-1]^{2}+\acute{V}(\varsigma )-\frac{1}{4}$, and motion equations
\begin{equation*}
\lbrack (\varsigma ^{\bullet })^{2}-1]\varsigma ^{\bullet \bullet }=-\frac{%
\partial \acute{V}}{\partial \varsigma }.
\end{equation*}%
Such equations were introduced in \cite{shapere17} for an example of
1-dTC-structure in condensed matter physics. The equations (\ref{1tqc})
provide a generalization for 1-TQCs modeled on a curved space-time.

A Lagrange density of type $\acute{L}[\varsigma (t)]$ (\ref{1tqclh}) may
describe a simplest example of 1-dTC-structure in cosmology theories if such
a value can be obtained effectively from a $\acute{L}[\varsigma
(x^{i},y^{a})]$ (\ref{1tqclg}) related to certain gravitational Lagrange
densities (for instance, describing dark energy, DE, configurations) and/or
matter sources (for corresponding dark matter, DM, models). Such
cosmological TC-structures are defined by some exact or parametric solutions
of gravitational field equations in a MGT, or in GR. In general, such
solutions are for metrics and (effective) matter fields depending both on
time and space coordinates which allows us to describe more realistically
inhomogeneous and/or locally anisotropic interactions of gravitational, DM
and standard matter fields. We shall consider details on elaborating such
cosmological models in next sections. Here, we note that 1-dTC-structures
with time depending $\acute{L}[\varsigma (t)]$ (\ref{1tqclh}) can be
extracted by additional assumptions/ constraints from certain classes of
generic off-diagonal solutions involving a $\acute{L}[\varsigma
(x^{i},y^{a})]$ (\ref{1tqclg}). We loose the possibility to generate
1-dTC-structures in general forms if we work from the very beginning only
with $\acute{L}[\varsigma (t)]$ (before constructing a class of solutions
with metrics and fields depending on $(x^{i},y^{a})$). This is the property
of nonlinear dynamics and respective systems of nonlinear partial
differential equations PDEs (see below formulas (\ref{mfeq}) and discussion
therein).

\subsubsection{Two dimensional relativistic time quasicrystal structures}

We can consider also a 2-d time quasicrystal structure, 2-TQC, defined by
two coupled scalar fields $\chi (x^{i},y^{a})$ and $\breve{q}(x^{i},y^{a}),$
for a model with effective planar charge in certain effective external
fields on a curved spacetime. The Lagrangian is postulated
\begin{equation}
\breve{L}(\chi (x^{i},y^{a}),q(x^{i},y^{a}))=\frac{\chi _{0}}{4}(\mathbf{g}%
^{\alpha \beta }(\mathbf{e}_{\alpha }\chi )(\mathbf{e}_{\beta }\chi ))^{2}+%
\breve{B}^{\gamma }(\chi )\mathbf{e}_{\gamma }\breve{q}-\breve{Q}(\chi )-%
\breve{V}(\breve{q}),  \label{2tqclg}
\end{equation}%
where the constant $\chi _{0},$ functionals $B^{\gamma }(\chi )$ and $Q(\chi
),$ and effective potential $\breve{V}(\breve{q})$ define a corresponding
time quasiperiodic structure. The resulting variotional motion equations are
\begin{eqnarray}
\chi _{0}\widehat{\mathbf{D}}^{\gamma }\widehat{\mathbf{D}}_{\gamma }\chi
&=&-\frac{\partial \breve{B}^{\gamma }(\chi )}{\partial \chi }\mathbf{e}%
_{\gamma }\breve{q}+\frac{\partial \breve{Q}(\chi )}{\partial \chi }  \notag
\\
\frac{\partial \breve{B}^{\gamma }(\chi )}{\partial \chi }\mathbf{e}_{\gamma
}\chi &=&-\frac{\partial \breve{V}}{\partial \breve{q}}.  \label{2tqc}
\end{eqnarray}

In a non-relativistic limit with time depending scalar fields $\chi (t)$ and
$q(t),$ we obtain%
\begin{equation}
\breve{L}(\chi (x^{i},y^{a}),q(x^{i},y^{a}))\rightarrow \breve{L}(\chi
(t),q(t))=\frac{\chi _{0}}{2}(\chi ^{\bullet })^{2}+\breve{B}^{4}(\chi
)q^{\bullet }-\breve{Q}(\chi )-\breve{V}(\breve{q})  \label{2tqclh}
\end{equation}%
and the motion equations
\begin{eqnarray*}
\chi _{0}\chi ^{\bullet \bullet } &=&\frac{\partial \breve{B}(\chi )}{%
\partial \chi }\breve{q}^{\bullet }-\frac{\partial \breve{Q}(\chi )}{%
\partial \chi } \\
\chi ^{\bullet }\frac{\partial \breve{B}(\chi )}{\partial \chi } &=&-\frac{%
\partial \breve{V}}{\partial \breve{q}},
\end{eqnarray*}%
which describe an effective dynamics considered to 2-d TC in \cite{shapere17}%
. For additional constraints, such an effective 2-d dynamics reproduces 1-d
TC dynamics with an effective small positive mass $\chi _{0}$ acting as a
regulator for the 1-d time crystal dynamical system.

In this work, our goal is to provide a relativistic curve space-time
generalization of the so-called sisyphus dynamics with microstructure and
ratcheting is possible if we consider effective Lagrange densities $\breve{L}%
(\chi (x^{i},y^{a}),q(x^{i},y^{a}))$ (\ref{2tqclg}) for 2-TQC structures.
Certain such configurations contain as particular cases 1-dTC-structure but
the extensions of nonlinear dynamics from lower to higher dimensions is not
trivial. For instance, we can use such $\breve{L}$ to model DM contributions
via an effective energy momentum tensor $\overline{\mathbf{T}}_{\alpha \beta
}$ computed for $\overline{\mathcal{L}}=\breve{L},$ see below formula (\ref%
{efstemt}). Exact and parametric cosmological solutions with DM 2-TQC
sources can be constructed for generalized Einstein equations (see below the
formulas (\ref{mfeq}) and, for respective energy-momentum tensor, (\ref%
{msourc})) as it is described in section \ref{s3}. We can model cosmological
2-TQC structures depending only on a time like coordinate by considering
certain limits/ constraints when $\breve{L}\rightarrow \breve{L}(\chi
(t),q(t))$ (\ref{2tqclh}). Finally, it should be emphasized that we can
generate pure gravitational 2-TQC structures if a $\breve{L}$ is used for
constructing a gravitational Lagrange density in a MGT or GR. This can be
used for modelling nonlinear vacuum gravitational interactions and/or DE
effects, see below discussions related to formula (\ref{stqcp}).

\subsubsection{Three dimensional QC structures on curved spaces}

Quasiperiodic structures can be modeled as spacelike configurations on
curved spacetimes (called also quasicrystals, QG), see details in \cite%
{vbubuianu17,amaral17}. Let us consider a 3+1 decomposition with space like
coordinates $x^{\grave{\imath}}$ (for $\grave{\imath}=1,2,3$), time like
coordinate $y^{4}=t,$ which is adapted to another 2+2 decomposition with
fibration by 3-d hypersurfaces $\widehat{\Xi }_{t}.$ We define a canonically
nonholonomically deformed Laplace operator $\ ^{b}\widehat{\Delta }:=(\ ^{b}%
\widehat{D})^{2}=b^{\grave{\imath}\grave{j}}\widehat{D}_{\grave{\imath}}%
\widehat{D}_{\grave{j}},$ for the 3-d part of a d-metric. The distortion of $%
\ ^{b}\Delta :=(\ ^{b}\nabla )^{2}$ can be defined on any $\widehat{\Xi }%
_{t} $ using a d-metric (\ref{odans}) and respective restrictions of $%
\widehat{\mathbf{D}}.$

For instance, a QC structure can be defined by a scalar field $\psi
(x^{i},y^{a})$ which is a solution of an evolution equation with conserved
dynamics of type
\begin{equation}
\psi ^{\bullet }=\ ^{b}\widehat{\Delta }\left[ \frac{\delta \breve{F}}{%
\delta \psi }\right] =-\ ^{b}\widehat{\Delta }(\Theta \psi +Q\psi ^{2}-\psi
^{3}),  \label{qcevoleq}
\end{equation}%
\ The functional $\breve{F}$ in (\ref{qcevoleq}) defines an effective free
energy
\begin{equation*}
\breve{F}[\psi ]=\int \left[ -\frac{1}{2}\psi \Theta \psi -\frac{Q}{3}\psi
^{3}+\frac{1}{4}\psi ^{4}\right] \sqrt{b}dx^{1}dx^{2}\delta y^{3},
\end{equation*}%
where $b=\det |b_{\grave{\imath}\grave{j}}|,\delta y^{3}=\mathbf{e}^{3}$ and
the operator $\Theta $ and parameter $Q$ are defined in the partner work
\cite{vbubuianu17,amaral17}. Values $\Theta $ and $Q$ determine certain
quasiperiodic, aperiodic and/or QC order on curved spacetimes.

In this work, we show that cosmological solutions with QC structure can be
extended to more general classes with TQC. This is possible if we generalize
functionals $\breve{F}$ in (\ref{qcevoleq}) in some forms including
additional dependencies, for instance, on effective Lagrange densities $%
\breve{L}$ (\ref{2tqclg}) for 2-TQC structures. In next subsection we
consider an explicit such example.

\subsubsection{Mixed 2-d TQC and 3-d QC configurations}

We can model on $(\mathbf{V,g,N)}$ nonholonomic configurations both with TQC
and QC structure $(\chi ,q).$ Such a model can be generated by a function $%
\chi $ as in (\ref{2tqc}) when a function $q$ is additionally subjected to a
QC-evolution condition, when
\begin{eqnarray}
q^{\bullet } &=&\ ^{b}\widehat{\Delta }\left[ \frac{\delta \acute{F}}{\delta
q}\right] =-\ ^{b}\widehat{\Delta }(\acute{\Theta}q+\acute{Q}q^{2}-q^{3})%
\mbox{ for }  \notag \\
\acute{F}[\chi ,q] &=&\int \left[ -\frac{1}{2}q~\acute{\Theta}q-\frac{\acute{%
Q}}{3}q^{3}+\frac{1}{4}q^{4}\right] \sqrt{b}dx^{1}dx^{2}\delta y^{3}.
\label{qevoleq}
\end{eqnarray}%
In principle, we can generate more complex STQC structures when certain QC
conditions are imposed additionally for $\chi .$ Here we note that we wrote $%
\breve{F}$ in (\ref{qcevoleq}) and $\acute{F}$ (\ref{qevoleq}) in order to
avoid future ambiguities with conventions for MGT with $\mathbf{F}(\widehat{%
\mathbf{R}}),$ see below formula (\ref{fsourc}).

The scalar fields $(\chi ,q)$ subjected to the conditions of evolution (\ref%
{qevoleq}) can be included into certain effective Lagrange densities $\breve{%
L}$ (\ref{2tqclg}) and respective energy momentum tensor $\overline{\mathbf{T%
}}_{\alpha \beta }$ computed for $\ \ \overline{\mathcal{L}}=\breve{L}$ (\ref%
{efstemt}). In result, we can elaborate on cosmological models with DM and
DE evolution encoding both TQC and QC structures, see sections \ref{s4} and %
\ref{s5}.

\subsubsection{Matter sources and STQC structures in curved space-times}

There are two possibilities to generate STQC structures in gravity theories:

\paragraph{1. When some coefficients of metrics and/or (non) linear
connections with functional dependencies on STQC generating functions:}

Such configurations will be defined by certain data $\varsigma =$ $\overline{%
\varsigma }(x^{i},y^{a}),$ see (\ref{1tqc}); $\chi =\overline{\chi }%
(x^{i},y^{a})$ and $\breve{q}=\overline{\breve{q}}(x^{i},y^{a}),$ see (\ref%
{2tqc}); $\psi =\overline{\psi }(x^{i},y^{a}),$ see (\ref{qcevoleq}); and/or
$\chi =\overline{\chi }(x^{i},y^{a})$ and $q=\overline{q}(x^{i},y^{a}),$ see
(\ref{qevoleq}), contained in a functional form for a d-metric (\ref{odans})
\begin{equation}
\mathbf{g}_{\alpha \beta }=\overline{\mathbf{g}}_{\alpha \beta }[x^{i},y^{a};%
\overline{\varsigma },\overline{\chi },\overline{\breve{q}},\overline{\psi },%
\overline{q},...],  \label{stqcp}
\end{equation}%
where symbols are overlined in order to emphasize that such values are
considered for geometric objects defining tensor and/or connection fields in
a MGT or GR. In result, functional dependencies on $\overline{\varsigma },%
\overline{\chi },\overline{\breve{q}},\overline{q},...$(dots are used for
any other possible sets of generating functions, for instance, solitonic
waves, other QC structures etc) can be computed for (non) linear
connections, $N_{i}^{a}=\overline{N}_{i}^{a}[x^{i},y^{a};\overline{\varsigma
},\overline{\chi },\overline{\breve{q}},\overline{\psi },\overline{q},...]$
and $\mathbf{\Gamma }_{\ \beta \gamma }^{\alpha }=\overline{\mathbf{\Gamma }}%
_{\ \beta \gamma }^{\alpha }[x^{i},y^{a};\overline{\varsigma },\overline{%
\chi },\overline{\breve{q}},\overline{\psi },\overline{q},...]$ ; respective
curvatures, torsions etc. computed, for instance, for $\mathbf{\nabla
\lbrack \overline{\mathbf{g}}]}$ and/or $\widehat{\mathbf{D}}\mathbf{[%
\overline{\mathbf{g}}]}.$

In a series of our previous works \cite{cosmv2,cosmv3,vbubuianu17} and
references therein, we studied cosmological and black hole metrics which are
similar to (\ref{stqcp}) if they are not stationary but extended to
configurations with evolution and encoding possible QC structures. The main
goal of this article is prove that the AFDM allows constructions with STQC
structures for elaborating more realistic cosmological DE and DM models. In
explicit form, we consider examples with 1-TQC and 2-TQC configurations.

\paragraph{2. By matter fields as STQC sources:}

For simplicity, we can consider actions
\begin{equation}
~^{m}\mathcal{S}=\int d^{4}u\sqrt{|\mathbf{g}|}~^{m}\mathcal{L}  \label{actm}
\end{equation}%
for matter field Lagrange densities $~^{m}\mathcal{L}\left( \phi \right) $
[we label by $\phi (x^{i},y^{a})$ all necessary sets of matter fields from
standard particle physics and GR] depending only on coefficients of a metric
field and do not depend on their derivatives when
\begin{equation}
\ ^{m}\mathbf{T}_{\alpha \beta }:=-\frac{2}{\sqrt{|\mathbf{g}_{\mu \nu }|}}%
\frac{\delta (\sqrt{|\mathbf{g}_{\mu \nu }|}\ \ ^{m}\mathcal{L})}{\delta
\mathbf{g}^{\alpha \beta }}=\ ^{m}\mathcal{L}\mathbf{g}^{\alpha \beta }+2%
\frac{\delta (\ ^{m}\mathcal{L})}{\delta \mathbf{g}_{\alpha \beta }}.
\label{emtm}
\end{equation}

Additionally, we can consider Lagrange densities for 1-d, 2-d TQC, with
additional prescriptions for QC like structures as we used above for
deriving the motion equations (\ref{1tqc}), (\ref{2tqc}), and/or (\ref%
{qevoleq}). The sum of such Lagrange densities is written%
\begin{equation}
\overline{\mathcal{L}}[\varsigma ,\chi ,q,\psi ,\breve{q},...]=\acute{L}%
(\varsigma )+\breve{L}(\chi ,q)+...  \label{sumlagd}
\end{equation}%
which allows us to compute the energy-momentum tensor for STQC-matter,
\begin{equation}
\overline{\mathbf{T}}_{\alpha \beta }:=-\frac{2}{\sqrt{|\mathbf{g}_{\mu \nu
}|}}\frac{\delta (\sqrt{|\mathbf{g}_{\mu \nu }|}\ \ \overline{\mathcal{L}})}{%
\delta \mathbf{g}^{\alpha \beta }}.  \label{efstemt}
\end{equation}%
A source $~\overline{\mathbf{\Upsilon }}_{\mu \nu }=M_{P}^{-2}(\overline{%
\mathbf{T}}_{\mu \nu }-\frac{1}{2}\mathbf{g}_{\mu \nu }\overline{\mathbf{T}}%
_{\mu \nu }),$ with $\overline{\mathbf{T}}=\overline{\mathbf{T}}_{\mu \nu }%
\mathbf{g}^{\mu \nu }$ and $M_{P}$ being the Planck mass determined by the
gravitational constant, can be used for modelling DM effects with pattern
forming and STQC-structure.

Let us clarify the link between (effective) Lagrangians and functionals for
STQC structures considered in this section and the rest of the paper:
Observational data for modern accelerating cosmology and DE and DM physics
provide evidence that our Universe has a very complex nonlinear evolution
structure. For different space and time scales, the meta galactic structure
and dynamic can be conventionally described by certain STQC and/or QC
configurations with quasi periodic inhomogeneities and anisotropies,
nonlinear fluctuations etc. In realistic forms, such configurations can be
modelled by effective dynamical and evolution theories determined by sums of
Lagrange densities $\overline{\mathcal{L}}=\acute{L}(\varsigma )+\breve{L}%
(\chi ,q)+...$ (\ref{sumlagd}) including, for instance, terms for
generalized 1- and/or 2-TQC structures, see formulas (\ref{1tqclg}) and/or (%
\ref{2tqclg}).

The goal of this work is to show that cosmological models with STQC
structure can be defined by exact or parametric solutions of the
gravitational field equations in MGT and/or GR. In such an approach, the
effective fields modelling STQC configurations with a $\overline{\mathcal{L}}
$ have to be encoded into a metric structure $\overline{\mathbf{g}}_{\alpha
\beta }[x^{i},y^{a};\overline{\varsigma },\overline{\chi },\overline{\breve{q%
}},\overline{\psi },\overline{q},...]$ (\ref{stqcp}), and related (non)
linear connections, and/or an energy-momentum tensor for STQC-matter $%
\overline{\mathbf{T}}_{\alpha \beta }$ (\ref{efstemt}) as we explained
above. We develop also a geometric method for generating solutions of
(modified) Einstein equations which model gravitational and (effective)
matter field interactions with TC like and generalized STQC structure, see
reviews of results in \cite{cosmv2,cosmv3,vbubuianu17}. In our previous
works \cite{amaral17,vbubuianu17}, quasiperiodic configurations where
studied for space like QC structures. In section \ref{s3}, we show that the
anholonomic frame deformation method, ADFM, provides also an analytic
techniques for generating STQC cosmological solutions. There are used new
classes of generic off-diagonal ansatz, and considered their
diagonalization, in section \ref{ssparamcm}.

Examples of STQC, time QC and TC solutions are provided in explicit form in
section \ref{ssexamplsol}. The simplest (toy) example involving cosmological
solutions with 1-TQC structure when the energy momentum tensor $\overline{%
\mathbf{T}}_{\alpha \beta }[\acute{L}(\varsigma )]$ (\ref{efstemt})
modelling DM is computed for $\overline{\mathcal{L}}=\acute{L}(\varsigma )$ (%
\ref{1tqclg}) is analyzed in brief in subsection \ref{ssexamplsoltoy}. Here
we note that even such very simplified nonlinear evolution cosmological
models with 1-TQC are described by quite sophisticate formulas because time
and space like quasiperiodicity involves more general classes of generating
functions and effective scalar fields.

Another very important relations of (effective) Lagrangians and functionals
for time and space like (quasi) periodic configurations are stated in
section \ref{s4} \ (contributing substantially in different models of STQC)
and \ref{s5} (a late STQC cosmological dynamics with DE and DM
interactions). We prove that in all such (modified) gravitational theories,
effective Lagrange densities of type $\acute{L}(\varsigma )$ (\ref{1tqclg})
and $\breve{L}(\chi ,q)$ (\ref{2tqclg}) and functionals of type $\breve{F}%
[\psi ]$ and $\acute{F}[\chi ,q],$ see respective formulas (\ref{qcevoleq})
and (\ref{qevoleq}), describe more realistically the cosmological structure
and evolution of our Universe.

\subsection{Gravitational and matter field equations for STQCs}

The motion equations for a MGT with a functional $\mathbf{F}(\widehat{%
\mathbf{R}})$ for the Lagrange density for gravitational field can be
derived by a N--adapted variational calculus, see details in \cite%
{cosmv2,eegrg,cosmv3,rajpvcosm,amaral17,vbubuianu17} and references therein.%
\footnote{%
The geometric construction are similar to those for $f(R)$ gravity and
various generalizations \cite{capoz,nojod1,sami,nojiri,vagnozzi} but
generalized for nonholonomic manifolds with N-connection splitting (\ref%
{ncon}).} Such a system of nonlinear PDEs can be represented in an effective
Einstein form,
\begin{equation}
\widehat{\mathbf{R}}_{\mu \nu }=\mathbf{\Upsilon }_{\mu \nu }.  \label{mfeq}
\end{equation}%
with sources in the right side parameterized in the form
\begin{equation}
\mathbf{\Upsilon }_{\mu \nu }=~^{F}\mathbf{\Upsilon }_{\mu \nu }+~^{m}%
\mathbf{\Upsilon }_{\mu \nu }+~\overline{\mathbf{\Upsilon }}_{\mu \nu }.
\label{sourcc}
\end{equation}%
In these formulas, the effective energy-momentum tensor
\begin{equation}
\ \ ^{F}\mathbf{\Upsilon }_{\mu \nu }=(\frac{\mathbf{F}}{2~\mathbf{F}%
^{\prime }}-\frac{\widehat{\mathbf{D}}^{2}\mathbf{F}^{\prime }}{\mathbf{F}%
^{\prime }})\mathbf{g}_{\mu \nu }+\frac{\widehat{\mathbf{D}}_{\mu }\widehat{%
\mathbf{D}}_{\nu }\mathbf{F}^{\prime }}{\mathbf{F}^{\prime }}  \label{fsourc}
\end{equation}%
is determined by the functional $\mathbf{F}(\widehat{\mathbf{R}})$
determines an effective energy-momentum tensor.

We can always define nonholonomic variable for a large class of commutative
and noncommutative MGTs (including contributions, for superstring and/or
supergravity models, metric-affine gravity with nonmetricity and torsion
etc) when the modified (generalized) gravitational field equations can be
represented in the form (\ref{mfeq}). Corresponding effective and matter
field sources (\ref{sourcc}) may encode various contributions from extra
dimensions, distortions of connection structures, super symmetric and
noncommutative terms etc. For simplicity, in this work we consider only
effective sources $\ \ ^{F}\mathbf{\Upsilon }_{\mu \nu }$ (\ref{mfeq})
determined by $f(R)$--theories but generalized to distortions of linear
connections of type $\widehat{\mathbf{D}}=\nabla +\widehat{\mathbf{Z}},$
with a distortion d-tensor $\widehat{\mathbf{Z}}$ uniquely defined by
formula (\ref{twocon}). As we proved in our above cited works on the AFDM
and applications, it is convenient to use the \ the canonical d-connection $%
\widehat{\mathbf{D}}$ when formula written for a special class of N--adapted
frames allow certain general decoupling and integration of generalized
Einstein equations (\ref{mfeq}). Such general decoupling with generic
off-diagonal metrics depending, in principle, on all spacetime coordinates
via certain classes of generating and integration functions and generating
sources is not possible if we work only with the Levi-Civita connection $%
\nabla .$

The source for the matter fields and STQC matter configurations (they can be
for standard matter fields in GR, or for certain DM models, with
supgravity/noncommutative additional terms etc.) can be computed in standard
form,
\begin{equation}
~^{m}\mathbf{\Upsilon }_{\mu \nu }=\frac{1}{2M_{P}^{2}}\ ^{m}\mathbf{T}_{\mu
\nu }\mbox{ amd }\overline{\mathbf{\Upsilon }}_{\mu \nu }=\frac{1}{2M_{P}^{2}%
}\ \overline{\mathbf{T}}_{\mu \nu }.  \label{msourc}
\end{equation}%
Such (effective) energy-momentum tensors can be postulated in a geometric
form like in GR (but for generalized connections) and/or derived
equivalently following a N-adapted gravitationsl calculus. They are included
as respective terms for a generalized source $\mathbf{\Upsilon }_{\mu \nu }$
(\ref{sourcc}), when $\overline{\mathbf{\Upsilon }}_{\mu \nu }$ is
determined by the energy-momentum of certain STQC like matter fields. Such
quasiperiodic time and space structures can be studied by the same geometric
methods both in GR and MGTs. In this work we consider generalized
constructions for $\mathbf{F}(\widehat{\mathbf{R}})$-gravity because such
theories are intensively studied for elaborating models of acclerating
cosmology and DE and DM theories. It is also important to show that the AFDM
can be applied for generating solutions with STQC structure in the bulk of
modern gravity theories.

In our previous works \cite{cosmv2,eegrg,cosmv3,vbubuianu17}, we proved that
we can construct exact solutions for the system (\ref{mfeq}) in explicit
form for any source (\ref{sourcc}) which via frame transforms $\mathbf{%
\Upsilon }_{\mu \nu }=e_{\ \mu }^{\mu ^{\prime }}e_{\ \nu }^{\nu ^{\prime }}%
\mathbf{\Upsilon }_{\mu ^{\prime }\nu ^{\prime }}$ can be parameterized into
N--adapted diagonalized form
\begin{equation}
\mathbf{\Upsilon }_{\ \nu }^{\mu }=diag[\ ~_{h}\mathbf{\Upsilon }%
(x^{i}),~_{h}\mathbf{\Upsilon (}x^{i}),\mathbf{\Upsilon }(x^{i},t),~\mathbf{%
\Upsilon }(x^{i},t)].  \label{sources}
\end{equation}%
In these formulas, the values
\begin{equation}
~_{h}\Upsilon (x^{i})=\ ~_{h}^{F}\Upsilon (x^{i})+\ ~_{h}^{m}\Upsilon
(x^{i})+\ ~_{h}\overline{\Upsilon }(x^{i})\mbox{ and }\Upsilon (x^{i},t)=\
~^{F}\Upsilon (x^{i},t)+\ ^{m}\Upsilon (x^{i},t)+\ \overline{\Upsilon }%
(x^{i},t)  \label{sourcparam}
\end{equation}%
are considered as generating source functions which impose certain
nonholonomic constraints on the dynamics of (effective) matter fields.
Having constructed a class of exact/parametric solutions we have to
prescribe (\ref{sourcparam}) in some forms which an explicit solution will
describe/explain observational/experimental data.\footnote{%
In principle, we can generate exact solutions for parameterizations of
metrics and sources depending on all space-time coordinates but formulas for
such solution are much more cumbersome. For our purposes, it is enough to
consider effective source configurations of type (\ref{sourcparam}).}

Finally, we note that the Einstein equations for GR are obtained as a
particular case if $\ \mathbf{F}(\widehat{\mathbf{R}})=R$ with a
parametrization when $\ ^{F}\mathbf{\Upsilon }_{\mu \nu }[R]=0$ for $%
\widehat{\mathbf{D}}_{\mid \widehat{\mathcal{T}}\rightarrow 0}=\mathbf{%
\nabla }$. Nontrivial STQC sources $\overline{\mathbf{\Upsilon }}_{\ \nu
}^{\mu }=diag[~_{h}\overline{\mathbf{\Upsilon }}(x^{i}),~_{h}\overline{%
\mathbf{\Upsilon }}(x^{i}),\ \overline{\mathbf{\Upsilon }}(x^{i},t),~%
\overline{\mathbf{\Upsilon }}(x^{i},t)]$ can be considered also in GR in
order to construct inflation scenarios and dark energy and dark matter
models.

\subsection{Parameterizations for cosmological d-metrics}

\label{ssparamcm}In this subsection, there are provided necessary
coefficient formulas and examples of N-adapted calculus used in locally
anisotropic cosmology with STQC structure. We consider certain basic
representations for quadratic line elements describing nonholonomic
deformations of prime metrics into target cosmological ones.

\subsubsection{Target d-metrics with polarization functions}

Let us consider a prime metric $\ \mathbf{\mathring{g}}=\mathring{g}_{\alpha
\beta }(x^{i},y^{a})du^{\alpha }\otimes du^{\beta }$ which can be written in
a N-adapted form
\begin{eqnarray}
\mathbf{\mathring{g}} &=&\mathring{g}_{\alpha }(u)\mathbf{\mathring{e}}%
^{\alpha }\otimes \mathbf{\mathring{e}}^{\beta }=\mathring{g}%
_{i}(x,y)dx^{i}\otimes dx^{i}+\mathring{h}_{a}(x,y)\mathbf{\mathring{e}}%
^{a}\otimes \mathbf{\mathring{e}}^{a},\mbox{ for }  \label{primedm} \\
&&\mathbf{\mathring{e}}^{\alpha }=(dx^{i},\mathbf{e}^{a}=dy^{a}+\mathring{N}%
_{i}^{a}(u)dx^{i}),\mbox{ and }\mathbf{\mathring{e}}_{\alpha }=(\mathbf{%
\mathring{e}}_{i}=\partial /\partial y^{a}-\mathring{N}_{i}^{b}(u)\partial
/\partial y^{b},\ {e}_{a}=\partial /\partial y^{a}).  \notag
\end{eqnarray}%
In general, such a d-metric can be, or not, a solution of some gravitational
field equations in a MGT or GR. We can take $\mathbf{\mathring{g}}$ (\ref%
{primedm}) as a necessary type coordinate transform of a cosmological
metric, for instance, of a Friedman--Lema\^{\i}tre--Robertson--Walker
(FLRW), or a Bianch type, metric. For a diagonalizable FLRW metric, we can
always find a frame/coordinate system when $\mathring{N}_{i}^{b}=0.$ It is
convenient to work with local coordinate systems and nonzero values for $%
\mathring{N}_{i}^{b}$ which do not result in certain singular/pecular
nonholonomic deformations as we shall describe below.

To generate new classes of cosmological solutions we shall consider
nonholonomic deformations $\mathbf{\mathring{g}}\rightarrow \widehat{\mathbf{%
g}}=[g_{\alpha }=\eta _{\alpha }\mathring{g}_{\alpha },N_{i}^{a}=\ \eta
_{i}^{a}\mathring{N}_{i}^{a}]$ with so-called $\eta $-polarization functions
of a 'prime' metric, $\mathbf{\mathring{g}}$, into a 'target' metric $%
\mathbf{g}=\widehat{\mathbf{g}}.$ In N-adapted form, target d-metrics of
type (\ref{odans}) are parameterized
\begin{eqnarray}
\ \mathbf{\mathring{g}} &\rightarrow &\widehat{\mathbf{g}}%
=g_{i}(x^{k})dx^{i}\otimes dx^{i}+h_{a}(x^{k},t)\mathbf{e}^{a}\otimes
\mathbf{e}^{a}  \label{dme} \\
&=&\eta _{i}(x^{k},y^{b})\mathring{g}_{i}dx^{i}\otimes dx^{i}+\eta
_{a}(x^{k},y^{b})\mathring{h}_{a}\mathbf{e}^{a}[\eta ]\otimes \mathbf{e}%
^{a}[\eta ],  \notag
\end{eqnarray}%
where the N-elongated basis (\ref{nder}) is written for $N_{i}^{a}(u)=\eta
_{i}^{a}(x^{k},y^{b})\mathring{N}_{i}^{a}(x^{k},y^{b}),$ i.e. in the form%
\footnote{%
we do not consider summation on repeating indices if they are not written as
contraction of "up-low" ones} $\mathbf{e}^{\alpha }[\eta ]=(dx^{i},\mathbf{e}%
^{a}=dy^{a}+\eta _{i}^{a}\mathring{N}_{i}^{a}dx^{i}).$ We shall subject a $%
\widehat{\mathbf{g}}$ to the condition that it defines a generalized
cosmological solution of (modified) Einstein equations when coefficients of
a corresponding d-metric and (non) linear connection depend at least on a
time like coordinate $y^{4}=t.$

Target quadratic line elements can be represented in generic off-diagonal
form, \newline
$\mathbf{g}_{\alpha \beta }=[g_{i},h_{a},n_{i},w_{i}],$ and/or using $\eta $%
-polarization functions,
\begin{eqnarray}
ds^{2}
&=&g_{i}(x^{k})[dx^{i}]^{2}+h_{3}(x^{k},t)[dy^{3}+n_{i}(x^{k},t)dx^{i}]^{2}+
h_{4}(x^{k},t)[dt+w_{i}(x^{k},t)dx^{i}]^{2}  \label{targ1} \\
&=&\eta _{i}(x^{k},t)\mathring{g}_{i}(x^{k},t)[dx^{i}]^{2}+\eta _{3}(x^{k},t)%
\mathring{h}_{3}(x^{k},t)[dy^{3}+\eta _{i}^{3}(x^{k},t)\mathring{N}%
_{i}^{3}(x^{k},t)dx^{i}]^{2}  \notag \\
&&+\eta _{4}(x^{k},t)\mathring{h}_{4}(x^{k},t)[dt+\eta _{i}^{4}(x^{k},t)%
\mathring{N}_{i}^{4}(x^{k},t)dx^{i}]^{2}  \notag \\
&=&\eta _{i}\mathring{g}_{i}[dx^{i}]^{2}+\eta _{3}\mathring{h}%
_{3}[dy^{3}+\eta _{k}^{3}\mathring{N}_{k}^{3}dx^{k}]^{2}+\eta _{4}\mathring{h%
}_{4}[dt+\eta _{k}^{4}\mathring{N}_{k}^{4}dx^{k}]^{2}.  \label{targ3}
\end{eqnarray}%
Let us consider a coordinate transform to a new time like coordinate $%
y^{4}=t\rightarrow \tau $ when $t=t(x^{i},\tau ),$
\begin{eqnarray*}
dt &=&\partial _{i}tdx^{i}+(\partial t/\partial \tau )d\tau ;d\tau
=(\partial t/\partial \tau )^{-1}(dt-\partial _{i}tdx^{i}),\mbox{ i.e. } \\
&& (\partial t/\partial \tau )d\tau =(dt-\partial _{i}tdx^{i}).
\end{eqnarray*}%
We can rewrite the target d-metric using the new time variable $\tau .$ For
instance, the computations for the 4th term in (\ref{targ3}) are
\begin{eqnarray*}
&& \eta _{4}\mathring{h}_{4}[dt+\eta _{k}^{4}\mathring{N}_{k}^{4}dx^{k}]^{2}
=\eta _{4}\mathring{h}_{4}[\partial _{k}tdx^{k}+(\partial t/\partial \tau
)d\tau +\eta _{k}^{4}\mathring{N}_{k}^{4}dx^{k}]^{2} \\
&&=\eta _{4}\mathring{h}_{4}[(\partial _{k}t)dx^{k}+(\partial t/\partial
\tau )d\tau +\eta _{k}^{4}\mathring{N}_{k}^{4}dx^{k}]^{2} = \eta _{4}%
\mathring{h}_{4}[(\partial t/\partial \tau )d\tau +(\partial _{k}t+\eta
_{k}^{4}\mathring{N}_{k}^{4})dx^{k}]^{2} \\
&&=\mathring{h}_{4}[\eta _{4}(\partial t/\partial \tau )d\tau +\eta
_{4}(\partial _{k}t+\eta _{k}^{4}\mathring{N}_{k}^{4})dx^{k}]^{2} =
\mathring{h}_{4}[\eta _{4}(\partial t/\partial \tau )d\tau +\eta
_{4}(\partial _{k}t/\mathring{N}_{k}^{4}+\eta _{k}^{4})\mathring{N}%
_{k}^{4}dx^{k}]^{2}
\end{eqnarray*}%
Considering
\begin{eqnarray*}
\eta _{4}(\partial t/\partial \tau ) &=&1;\partial t/\partial \tau =(\eta
_{4})^{-1}\mbox{ introduced for } dt=\partial _{i}tdx^{i}+(\partial
t/\partial \tau )d\tau , \mbox{ when } \\
dt &=&\partial _{i}tdx^{i}+(\eta _{4})^{-1}d\tau \mbox{ for }\check{\eta}%
_{k}^{4}=\eta _{4}(\partial _{k}t/\mathring{N}_{k}^{4}+\eta _{k}^{4})%
\mathring{N}_{k}^{4},
\end{eqnarray*}%
a new time coordinate $\tau $ can be found from formulas $\partial
t/\partial \tau =(\eta _{4})^{-1}$ resulting in
\begin{equation*}
d\tau =\eta _{4}(x^{k},t)dt;\tau =\int \eta _{4}(x^{k},t)dt+\tau _{0}(x^{k}).
\end{equation*}%
Such coordinates with $\tau$ are useful for computations of nonholonomic
deformations of the FLRW metrics.

\subsubsection{Parameterizations of prime d-metrics}

Let us consider a target line quadratic element for an off-diagonal
cosmological solution in MGT written in the form (\ref{targ3}). We can
rewrite it using an effective target locally anisotropic cosmological
scaling factor $\check{a}^{2}(x^{k},\tau ):=\eta (x^{k},\tau )\mathring{a}%
^{2}(x^{i},\tau )$ with gravitational polarization $\eta (x^{k},\tau )$ and
prime cosmological scaling factor $\mathring{a}^{2}(x^{i},\tau ).$ This can
be performed following formulas%
\begin{eqnarray}
ds^{2} &=&\eta _{3}\{\frac{\eta _{i}}{\eta _{3}}\mathring{g}_{i}[dx^{i}]^{2}+%
\mathring{h}_{3}[dy^{3}+\eta _{k}^{3}\mathring{N}_{k}^{3}dx^{k}]^{2}\}+%
\mathring{h}_{4}[d\tau +\check{\eta}_{k}^{4}\mathring{N}_{k}^{4}dx^{k}]^{2}
\label{targ4} \\
&=&\check{a}^{2}(x^{k},\tau )\{\check{\eta}_{i}(x^{k},\tau )\mathring{g}%
_{i}[dx^{i}]^{2}+\mathring{h}_{3}[dy^{3}+\check{\eta}_{k}^{3}(x^{k},\tau )%
\mathring{N}_{k}^{3}dx^{k}]^{2}\}+\mathring{h}_{4}[d\tau +\check{\eta}%
_{k}^{4}(x^{k},\tau )\mathring{N}_{k}^{4}dx^{k}]^{2}  \notag \\
&=&\eta (x^{k},\tau )\mathring{a}^{2}(x^{i},\tau )\{\check{\eta}%
_{i}(x^{k},\tau )\mathring{g}_{i}[dx^{i}]^{2}+\mathring{h}_{3}[dy^{3}+\check{%
\eta}_{k}^{3}(x^{k},\tau )\mathring{N}_{k}^{3}dx^{k}]^{2}\}+\mathring{h}%
_{4}[d\tau +\check{\eta}_{k}^{4}(x^{k},\tau )\mathring{N}_{k}^{4}dx^{k}]^{2},
\notag
\end{eqnarray}%
\begin{eqnarray*}
\mbox{where }\check{a}^{2}(x^{k},\tau ):= &&\eta _{3}(x^{k},t(x^{i},\tau
))=\eta (x^{k},t(x^{i},\tau ))\mathring{a}^{2}(x^{k},t(x^{i},\tau ))=\eta
(x^{k},\tau )\mathring{a}^{2}(x^{i},\tau ); \\
\check{\eta}_{i}(x^{k},\tau ):= &&\frac{\eta _{i}(x^{k},t(x^{i},\tau ))}{%
\eta (x^{k},t(x^{i},\tau ))};\ \check{\eta}_{k}^{3}(x^{k},\tau ):=\eta
_{k}^{3}(x^{k},t(x^{i},\tau )); \\
\check{\eta}_{k}^{4}(x^{k},\tau ):= &&\eta _{4}\{\partial _{k}t(x^{i},\tau )[%
\mathring{N}_{k}^{4}(x^{i},t(x^{i},\tau ))]^{-1}+\eta
_{k}^{4}(x^{i},t(x^{i},\tau ))\}\mathring{N}_{k}^{4}(x^{i},t(x^{i},\tau ))
\end{eqnarray*}

If we consider a prime d-metric as a flat FLRW metric written in local
coordinates%
\begin{equation*}
\overline{u}=\{\overline{u}^{\alpha }(x^{i},y^{3},\tau )=(\overline{x}%
^{1}(x^{i},y^{3},\tau ),\overline{x}^{2}(x^{i},y^{3},\tau ),\overline{y}%
^{3}(x^{i},y^{3},\tau ),\overline{y}^{4}(x^{i},y^{3},\tau ))\},
\end{equation*}%
a d-metric (\ref{targ1}) can be written in curved coordinate form $\mathring{%
a}^{2}(\overline{u}),$ with $\overline{u}^{\alpha },$ and/or using a prime
cosmological scaling factor $\mathring{a}^{2}(\tau ),$
\begin{eqnarray}
d\mathring{s}^{2} &=&\mathring{a}^{2}(\overline{u})\{\mathring{g}_{i}(%
\overline{u})[d\overline{x}^{i}]^{2}+\mathring{h}_{3}(\overline{u})[d%
\overline{y}^{3}+\mathring{N}_{k}^{3}(\overline{u})d\overline{x}^{k}]^{2}\}+%
\mathring{h}_{4}(\overline{u})[d\overline{y}^{4}+\mathring{N}_{k}^{4}(%
\overline{u})d\overline{x}^{k}]^{2}  \notag \\
&\rightarrow &\mathring{a}^{2}(\tau )[dx^{\check{i}}]^{2}-d\tau ^{2},
\label{primedm1} \\
\mbox{ for }\overline{u}^{\alpha } &\rightarrow &(x^{i},y^{3},\tau ),%
\mathring{g}_{i}\rightarrow 1,\mathring{h}_{3}\rightarrow 1,\mathring{h}%
_{4}\rightarrow -1,\mathring{N}_{k}^{a}(\overline{u})\rightarrow 0%
\mbox{ and
}\mathring{a}^{2}(\overline{u})\rightarrow \mathring{a}^{2}(\tau ).  \notag
\end{eqnarray}%
Quasi FLRW configurations are determined (by definition) by a diagonalized
solution in MGT or GR for the d-metric is of type (\ref{targ4}) when the
integration functions and coordinates result in $\check{\eta}%
_{k}^{a}(x^{k},\tau )=0,$%
\begin{equation}
ds^{2}=\eta (x^{k},\tau )\mathring{a}^{2}\{\check{\eta}_{i}(x^{k},\tau )%
\mathring{g}_{i}[dx^{i}]^{2}+\mathring{h}_{3}[dy^{3}]^{2}\}+\mathring{h}%
_{4}[d\tau ]^{2}.  \label{qflrwm}
\end{equation}%
For partial small nonholonomic deformations, such d-metrics admit
parameterizations $\check{\eta}_{i}\simeq $ $1+\varepsilon \check{\chi}%
_{i}(x^{k},\tau )$ when the polarization of the target cosmological factor, $%
\eta (x^{k},\tau )$ can be a general one and not a value of type $%
1+\varepsilon \chi (x^{k},\tau )$ with a small parameter $\varepsilon .$ A
resulting scaling factor $a^{2}(x^{k},\tau )=\eta (x^{k},\tau )\mathring{a}%
^{2}(x^{k},\tau ),$ with possible further re-parametrization or a limit to $%
a^{2}(\tau )=\eta (\tau )\mathring{a}^{2}(\tau ),$ encodes possible
nonlinear off-diagonal and parametric interactions determined by systems of
nonlinear PDEs.

\subsubsection{Approximations for target d-metrics}

To study properties of MGTs and cosmological models is convenient to
consider different types of approximations for nonholonomic deformations of (%
\ref{primedm1}) to a target d-metric (\ref{targ4}). Let us analyze six
classes of exact, or parametric, solutions which can be generated by a
respective subclass of generating functions and/or generating sources, or
some diagonal approximations, or by introducing small $\varepsilon $%
-parameters.

\begin{enumerate}
\item We can chose mutual reparameterizations of generating functions $(\Psi
,\Upsilon )\iff (\Phi ,\Lambda =const)$ and integrating functions when the
coefficients of a target d-metric $\widehat{\mathbf{g}}_{\alpha \beta }(\tau
)$ depend only a time like coordinate $\tau ,$ when $\eta (x^{k},\tau
)\rightarrow $ $\widetilde{\eta }(\tau )$ and $a(x^{k},\tau )\rightarrow
\widetilde{a}^{2}(\tau )=$ $\widetilde{\eta }(\tau )\mathring{a}^{2}(\tau ).$
Respective linear quadratic elements (\ref{targ4}) can be represented in the
form
\begin{equation}
ds^{2}=\eta (\tau )\mathring{a}^{2}(\tau )\{\check{\eta}_{i}(\tau )\mathring{%
g}_{i}[dx^{i}]^{2}+\mathring{h}_{3}[dy^{3}+\check{\eta}_{k}^{3}(\tau )%
\mathring{N}_{k}^{3}dx^{k}]^{2}\}+\mathring{h}_{4}[d\tau +\check{\eta}%
_{k}^{4}(\tau )\mathring{N}_{k}^{4}dx^{k}]^{2}.  \label{targdm4a}
\end{equation}%
With respect to coordinate bases, such cosmological solutions can be generic
off-diagonal and could be chosen in some forms describing certain
nonholonomic deformations of Bianchi cosmological models.

\item For FLRW prime configurations, we can consider generation functions
and integration functions which result in zero values of the target
N-connection coefficients (\ref{nconnect}) and/or consider limits $\mathring{%
N}_{k}^{a}\rightarrow 0.$ For such cases, we transform (\ref{targdm4a}) into
diagonal metrics
\begin{equation}
ds^{2}=\eta (\tau )\mathring{a}^{2}(\tau )\{\check{\eta}_{i}(\tau )\mathring{%
g}_{i}[dx^{i}]^{2}+\mathring{h}_{3}(dy^{3})^{2}\}+\mathring{h}_{4}(d\tau
)^{2}  \label{targdm4b}
\end{equation}%
modeling certain locally anisotropic interactions with a "memory" of
nonholonomic and off-diagonal structures.

\item Small parametric nonholonomic deformations of a prime metric (\ref%
{primedm1}) into target off-diagonal cosmological solutions (\ref{targ4})
can be described by approximations
\begin{equation*}
\check{\eta}_{i}(x^{k},\tau )\simeq 1+\varepsilon _{i}\check{\chi}%
_{i}(x^{k},\tau ),\eta (x^{k},\tau )\simeq 1+\varepsilon _{3}\chi
(x^{k},\tau ),\check{\eta}_{k}^{a}(x^{k},\tau )\simeq 1+\varepsilon _{k}^{a}%
\check{\chi}_{k}^{a}(x^{k},\tau ),
\end{equation*}%
where small parameters $\varepsilon _{i},\varepsilon _{3},\varepsilon
_{k}^{a}$ satisfy conditions of type $0\leq |\varepsilon _{i}|,|\varepsilon
_{3}|,|\varepsilon _{k}^{a}|\ll 1$ and, for instance, $\chi (x^{k},\tau )$
is taken as a generating function. Such approximations may impose certain
relations between such $\varepsilon $-constants and $\chi $-functions and
restrict the classe of generating functions subjected to nonlinear
symmetries. Corresponding quadratic line elements are written%
\begin{eqnarray}
ds^{2} &=&[1+\varepsilon _{3}\chi (x^{k},\tau )]\mathring{a}^{2}(x^{i},\tau
)\{[1+\varepsilon _{i}\check{\chi}_{i}(x^{k},\tau )]\mathring{g}%
_{i}[dx^{i}]^{2}+  \label{targdm4c} \\
&&\mathring{h}_{3}[dy^{3}+(1+\varepsilon _{k}^{3}\check{\chi}%
_{k}^{3}(x^{k},\tau ))\mathring{N}_{k}^{3}dx^{k}]^{2}\}+\mathring{h}%
_{4}[d\tau +(1+\varepsilon _{k}^{4}\check{\chi}_{k}^{4}(x^{k},\tau ))%
\mathring{N}_{k}^{4}dx^{k}]^{2}.  \notag
\end{eqnarray}%
Such off-diagonal solutions define cosmological metrics with certain small
independent fluctuations, for instance, a FLRW embedded self-consistently
into a locally anisotropic

\item We can construct off-diagonal cosmological solutions with small
parameters $\varepsilon _{i},\varepsilon _{3},\varepsilon _{k}^{a}$ when the
generating functions and d-metric and N-connection coefficients do not
depend on space like coordinates. For such approximations, the quadratic
line element (\ref{targdm4c}) transforms into%
\begin{eqnarray}
ds^{2} &=&[1+\varepsilon _{3}\chi (\tau )]\mathring{a}^{2}(\tau
)\{[1+\varepsilon _{i}\check{\chi}_{i}(\tau )]\mathring{g}_{i}[dx^{i}]^{2}+
\label{targdm4d} \\
&&\mathring{h}_{3}[dy^{3}+(1+\varepsilon _{k}^{3}\check{\chi}_{k}^{3}(\tau ))%
\mathring{N}_{k}^{3}dx^{k}]^{2}\}+\mathring{h}_{4}[d\tau +(1+\varepsilon
_{k}^{4}\check{\chi}_{k}^{4}(\tau ))\mathring{N}_{k}^{4}dx^{k}]^{2}.  \notag
\end{eqnarray}

\item Off-diagonal deformations, for instance, of a FLRW metric into locally
anisotropic cosmological solutions in a MGT or GR can be constructed using
only one small parameter $\varepsilon =\varepsilon _{i}=\varepsilon
_{3}=\varepsilon _{k}^{a},$ when (\ref{targdm4d}) transforms into%
\begin{eqnarray}
ds^{2} &=&[1+\varepsilon \chi (x^{k},\tau )]\mathring{a}^{2}(x^{i},\tau
)\{[1+\varepsilon \check{\chi}_{i}(x^{k},\tau )]\mathring{g}_{i}[dx^{i}]^{2}+
\label{targdm4e} \\
&&\mathring{h}_{3}[dy^{3}+(1+\varepsilon \check{\chi}_{k}^{3}(x^{k},\tau ))%
\mathring{N}_{k}^{3}dx^{k}]^{2}\}+\mathring{h}_{4}[d\tau +(1+\varepsilon
\check{\chi}_{k}^{4}(x^{k},\tau ))\mathring{N}_{k}^{4}dx^{k}]^{2}.  \notag
\end{eqnarray}

\item Finally, we can impose on (\ref{targdm4e}) the condition that the $%
\varepsilon $--deformations depend on a time like coordinate. This results
in d-metrics
\begin{eqnarray}
ds^{2} &=&[1+\varepsilon \chi (\tau )]\mathring{a}^{2}(\tau
)\{[1+\varepsilon \check{\chi}_{i}(\tau )]\mathring{g}_{i}[dx^{i}]^{2}+
\label{targdm4f} \\
&&\mathring{h}_{3}[dy^{3}+(1+\varepsilon \check{\chi}_{k}^{3}(\tau ))%
\mathring{N}_{k}^{3}dx^{k}]^{2}\}+\mathring{h}_{4}[d\tau +(1+\varepsilon
\check{\chi}_{k}^{4}(\tau ))\mathring{N}_{k}^{4}dx^{k}]^{2}  \notag
\end{eqnarray}%
which can be considered as some ansatz used, for instance, for describing
quantum fluctuations of FLRW metrics in various MGTs.
\end{enumerate}

Various classes of cosmological solutions with parametric $\varepsilon $%
-decompositions can be performed in a self-consistent form by omitting
quadratic and higher terms after a class of solutions have been found for
some general data $(\eta _{\alpha },\eta _{i}^{a}).$ They are more general
than approximate solutions found, for instance, for classical and quantum
fluctuations of standard FLRW metrics. For certain subclasses of generic
off-diagonal solutions, we can consider that $\varepsilon _{i},\varepsilon
_{a},\varepsilon _{i}^{a}\sim \varepsilon ,$ when only one small parameter
is considered for all coefficients of nonholonomic deformations. This way,
we can generate various classes of nonholonomic small deformations of
cosmological solutions like in Refs. \cite%
{cosmv2,eegrg,cosmv3,rajpvcosm,amaral17,vbubuianu17}, see also references in
those works citing constructions on locally anisotropic black hole/ wormhole
and other type solutions.

\section{Off-diagonal and quasi FLRW solutions with STQC structure}

\label{s3}In this section, we show how the system of (modified) Einstein
equations can be integrated in general off-diagonal forms for locally
anisotropic and inhomogeneous cosmological solutions with one-Killing
symmetry and parameterized by metrics of type $\overline{\mathbf{g}}_{\alpha
\beta }[x^{i},y^{a};\overline{\varsigma },\overline{\chi},\overline{\breve{q}%
},\overline{\psi },\overline{q},...]$ (\ref{stqcp}). It is proven that after
certain classes of such generalized cosmological metrics and connections in
MGTs or GR have been found using geometric methods for integrating nonlinear
PDEs we can always extract quasi FLRW configurations encoding nonlinear
off-diagonal gravitational and matter field interactions. We emphasize that
cosmological STQC-solutions can not be constructed if we work from the very
beginning only with time depending metrics of type $\mathbf{g}_{\alpha \beta
}(t)$. Such particular type time like parameterizations, or approximations,
resulting in STQC-configurations must be considered at the end, when
generalized solutions with dependencies both on time and space coordinates
have been constructed. Such new classes of cosmological solutions will be
applied in next sections for elaborating new cosmological scenarios with
pattern forming and evolution of STQC-structures.

\subsection{Cosmological ansatz for STQC structures}

Any target d-metric (\ref{dme}) can be represented in a form (\ref{odans})
and/or in a quasi FLRW\ form [see details related to formula (\ref{qflrwm}%
)], with Killing symmetry on $\partial /\partial y^{3},$
\begin{eqnarray}
ds^{2}
&=&g_{i}(x^{k})[dx^{i}]^{2}+h_{3}(x^{k},t)[dy^{3}+n_{i}(x^{k},t)dx^{i}]^{2}+h_{4}(x^{k},t)[dt+w_{i}(x^{k},t)dx^{i}]^{2}
\label{dme1a} \\
&=&\eta _{i}\mathring{g}_{i}[\overline{e}^{i}]^{2}+\eta _{3}\mathring{h}_{3}[%
\overline{\mathbf{e}}^{3}]^{2}+\mathring{h}_{4}[\overline{\mathbf{e}}%
^{4}]^{2}  \label{targ3a} \\
&=&a^{2}(x^{k},\tau )\{\check{\eta}_{i}(x^{k},\tau )\mathring{g}_{i}[\check{e%
}^{i}]^{2}+\mathring{h}_{3}[\mathbf{\check{e}}^{3}]^{2}\}+\mathring{h}_{4}(%
\mathbf{\check{e}}^{4})^{2},  \label{targ4b} \\
\mbox{ where } &&\overline{e}^{i}=dx^{i},\overline{\mathbf{e}}%
^{3}=dy^{3}+\eta _{k}^{3}\mathring{N}_{k}^{3}dx^{k},\overline{\mathbf{e}}%
^{4}=\sqrt{|\eta _{4}|}(dt+\eta _{k}^{4}\mathring{N}_{k}^{4}dx^{k}),  \notag
\\
&&\check{e}^{i}=dx^{i},\mathbf{\check{e}}^{3}=dy^{3}+\check{\eta}%
_{k}^{3}(x^{k},\tau )\mathring{N}_{k}^{3}dx^{k},\mathbf{\check{e}}^{4}=d\tau
+\check{\eta}_{k}^{4}(x^{k},\tau )\mathring{N}_{k}^{4}dx^{k}.  \notag
\end{eqnarray}%
The formulas for frame and d-metric coefficient transforms in (\ref{dme1a}),
(\ref{targ3a}) and (\ref{targ4b}), and for certain parametric decompositions
and approximations, see (\ref{targ3}) and (\ref{targ4}), and (\ref{targdm4a}%
)-(\ref{targdm4f}). Here we note that the coordinate transforms $u^{\alpha
}=(x^{i},y^{3},y^{4}=t(x^{i},\tau ))\rightarrow (x^{i},y^{3},\tau )$ are
subjected to the conditions $d\tau =\eta _{4}(x^{k},t)dt$ and
\begin{eqnarray}
\sqrt{|h_{3}|} &=&\sqrt{|\eta _{3}\mathring{h}_{3}|}=a(x^{i},t(x^{i},\tau ))=%
\check{a}(x^{i},\tau )=\sqrt{|\eta (x^{i},\tau )|}\mathring{a}(x^{i},\tau ),
\label{coeffic} \\
&\simeq &a^{2}(x^{k},\tau )\rightarrow \widetilde{a}^{2}(\tau )=\widetilde{%
\eta }(\tau )\mathring{a}^{2}(\tau )%
\mbox{ for nonholonomic generation of
FLRW configurations. }  \notag
\end{eqnarray}%
These formulas can be considered for $\eta _{3}=\eta (x^{i},t)$ and $%
h_{3}=\eta _{3}\mathring{h}_{3}=a^{2}=\eta \mathring{a}^{2}$. If we choose
parameterizations and limits of solutinos with $\check{\eta}_{i}(\tau )=\eta
(\tau )=1,\mathring{g}_{i}=\mathring{h}_{3}=1$ and $\mathring{h}_{4}=-1,$
see (\ref{primedm1}), we obtain the flat FLRW metric used for elaborating
locally isotropic and homogeneous theories. For some nontrivial values, the
polarization functions $\eta (x^{i},\tau )$ or $\tilde{\eta}(\tau )$ define
respective nonholonomic deformations of some primary data $\mathring{a}%
(x^{i},t)$ or $\mathring{a}(t).$ Such $\eta $--functions can be found by
constructing generic off-diagonal solutions of gravitational field equations
in a MGT or GR.

\subsection{Systems of nonlinear PDEs for cosmological STQCs}

Off-diagonal cosmological solutions can be found in explicit form for ansatz
(\ref{dme1a}) if we apply the anholonomic frame deformation method, AFDM,
see details in Appendix A.2.2 of ref. \cite{vbubuianu17} (here we note that
we use a different system of notation in this work). Computing respective
Ricci d-tensors, the nonholonomic deformations of the Einstein equations (%
\ref{mfeq}) for the canonical d-connection $\widehat{\mathbf{D}}$ and
sources (\ref{sources}) can be written as a system of nonlinear PDEs,
\begin{eqnarray}
(\partial _{11}^{2}+\partial _{22}^{2})\ln |g_{1}| &=&2\ ~\ _{h}\Upsilon
\label{effeq} \\
(\ln |\Psi |)^{\bullet }\ h_{3}^{\bullet } &=&2h_{3}h_{4}\Upsilon  \notag \\
n_{i}^{\bullet \bullet }+\gamma n_{i}^{\bullet } &=&0,  \notag \\
\beta w_{i}-\alpha _{i} &=&0.  \notag
\end{eqnarray}%
The coefficients in these formulas are defined
\begin{eqnarray}
\alpha _{i} &=&h_{3}^{\bullet }\ \partial _{i}\ln |\Psi |,\ \beta
=h_{3}^{\bullet }\ (\ln |\Psi |)^{\bullet },\ \gamma =\left( \ln
|h_{3}|^{3/2}/h_{4}\right) ^{\bullet },  \notag \\
&&\mbox{ where }\Psi (x^{i},t)={2(\sqrt{|h_{3}|})^{\bullet }/\sqrt{|h_{4}|}},
\label{genf1b}
\end{eqnarray}%
for $h_{a}^{^{\bullet }}\neq 0$ and $(\ln |\Psi |)^{\bullet }\neq 0$ and $%
g_{1}(x^{k})=g_{2}(x^{k}).$ We use such systems of references and
coordinates when these conditions are satisfied.

The system (\ref{effeq}) can integrated for any generating function $h_{4}$
and sources $\ _{h}\Upsilon (x^{i})$ and $\Upsilon (x^{k},t).$
Alternatively, we can take $\Psi (x^{i},t)$ as a generating function and
then find the coefficients of the v-metric and N-connection.

\subsection{Generating functions and integrals for off-diagonal cosmological
solutions}

The system (\ref{effeq}) and (\ref{genf1b}) possess a property of decoupling
for unknown variables $g_{1},h_{a},$ $n_{i}$ and $w_{i},$ for a generating
function $\Psi ,$ see similar details in \cite%
{cosmv2,eegrg,cosmv3,vbubuianu17}. Integrating consequently, we generate
exact solutions for the modified Einstein equations (\ref{mfeq}). Here we
note an important nonlinear symmetry following from the second lines in (\ref%
{effeq}) \ and (\ref{genf1b}). Such equations can be written for the
cosmological scaling factor $a$,
\begin{eqnarray*}
(\ln |\Psi |)^{\bullet }\ a^{\bullet } &=&ah_{4}\Upsilon , \\
\Psi (x^{i},t) &=&{2}a{^{\bullet }/\sqrt{|h_{4}|}},
\end{eqnarray*}%
which allows us to redefine the generating functions and generating sources,
$(\Psi ,\Upsilon )\iff (\Phi ,\Lambda =const),$ subjected to the conditions
\begin{equation}
\frac{(\Psi ^{2})^{\bullet }}{|\Upsilon |}=\frac{(\Phi ^{2})^{\bullet }}{%
\Lambda },\mbox{ or  }\Lambda \ \Psi ^{2}=\Phi ^{2}|\Upsilon |-\int dt\ \Phi
^{2}|\Upsilon |^{\bullet }.  \label{nsym1b}
\end{equation}%
Using such formulas, we introduce a new generating function $\Phi ({x}%
^{i},t) $ and an (effective) cosmological constant $\Lambda \neq 0$ (for
zero values, there are necessary more special considerations), which can be
applied for generating exact off-diagonal solutions in explicit forms. The
value $\Lambda $ is fixed by observational data (for instance, for the dark
energy encoding contributions of nonlinear gravitational interactions with
nontrivial vacuum) or computed from classical/quantum models. We can
consider models with splitting of the cosmological constant into
contributions from modified gravity, matter fields and effective sources
from STQC structures, $\Lambda =\ ^{F}\Lambda +\ ^{m}\Lambda +\ \overline{%
\Lambda }.$

In N-adapted and/or equivalent off-diagonal forms, generic off-diagonal
solutions with Killing symmetry on $\partial /\partial y^{3}$ are determined
by such coefficients of d--metrics,
\begin{eqnarray}
\ g_{1} &=&g_{2}(x^{k})%
\mbox{ determined by a solution of 2-d Poisson eqs.
with source }2~(~_{h}^{F}\mathbf{\Upsilon }+~_{h}^{m}\mathbf{\Upsilon }%
+~_{~h}\overline{\mathbf{\Upsilon }});  \notag \\
h_{3} &=&a^{2}=h_{3}^{[0]}(x^{k})-\int dt\frac{(\Psi ^{2}){^{\bullet }}}{%
4\Upsilon }=h_{3}[\Phi ]=h_{3}^{[0]}(x^{k})-\frac{\Phi ^{2}}{4\Lambda }
\label{offdcosm} \\
&=&h_{3}^{[0]}(x^{k})-\int dt\frac{(\Psi ^{2}){^{\bullet }}}{4(~^{F}\mathbf{%
\Upsilon }+~^{m}\mathbf{\Upsilon }+~\overline{\mathbf{\Upsilon }})}%
=h_{3}^{[0]}(x^{k})-\frac{\Phi ^{2}}{4(\ ^{F}\Lambda +\ ^{m}\Lambda +\
\overline{\Lambda })};  \notag \\
h_{4} &=&-\frac{(\Psi ^{2}){^{\bullet }}}{4\Upsilon ^{2}h_{3}}=-\frac{(\Psi
^{2}){^{\bullet }}}{4(~^{F}\mathbf{\Upsilon }+~^{m}\mathbf{\Upsilon }+~%
\overline{\mathbf{\Upsilon }})^{2}h_{3}}  \notag \\
&=&-\frac{|(\Phi ^{2})^{{^{\bullet }}}|^{2}}{4h_{3}[\Phi ]|(\ ^{F}\Lambda +\
^{m}\Lambda +\ \overline{\Lambda })\int dt(~^{F}\mathbf{\Upsilon }+~^{m}%
\mathbf{\Upsilon }+~\overline{\mathbf{\Upsilon }})(\Phi ^{2})^{{^{\bullet }}%
}|};  \notag
\end{eqnarray}%
and the coefficients of N--connection,
\begin{eqnarray}
N_{k}^{3} &=&n_{k}({x}^{i},t)=\ \ _{1}n_{k}(x^{i})+\ _{2}n_{k}(x^{i})\int dt%
\frac{(\Phi {^{\bullet }})^{2}}{4|\Lambda \int dt~\mathbf{\Upsilon }(\Phi
^{2})^{{^{\bullet }}}|\ |h_{3}|^{5/2}}  \label{nconnect} \\
&=&\ _{1}n_{k}(x^{i})+\ _{2}n_{k}(x^{i})\int dt\frac{(\Phi {^{\bullet }})^{2}%
}{4|(\ ^{F}\Lambda +\ ^{m}\Lambda +\ \overline{\Lambda })(~^{F}\mathbf{%
\Upsilon }+~^{m}\mathbf{\Upsilon }+~\overline{\mathbf{\Upsilon }})(\Phi
^{2})^{{^{\bullet }}}|\ |h_{3}|^{5/2}};  \notag \\
N_{i}^{4} &=&w_{i}({x}^{i},t)=\frac{\partial _{i}\ \Psi }{\Psi {^{\bullet }}}%
=\frac{\partial _{i}[\int dt\ \Upsilon (\Phi ^{2}){^{\bullet }}]}{\Upsilon
(\Phi ^{2}){^{\bullet }}}=\frac{\partial _{i}[\int dt\ (~^{F}\mathbf{%
\Upsilon }+~^{m}\mathbf{\Upsilon }+~\overline{\mathbf{\Upsilon }})(\Phi ^{2})%
{^{\bullet }}]}{(~^{F}\mathbf{\Upsilon }+~^{m}\mathbf{\Upsilon }+~\overline{%
\mathbf{\Upsilon }})(\Phi ^{2}){^{\bullet }}}.\ \   \notag
\end{eqnarray}%
In above formulas, the values $h_{3}^{[0]}(x^{k}),$ $\ _{1}n_{k}(x^{i}),$
and $\ _{2}n_{k}(x^{i})$ are integration functions encoding various possible
sets of (non) commutative parameters and integration constants.

Any coefficient $h_{3}$ with $h_{3}^{\bullet }\neq 0,$ in (\ref{offdcosm})
can be considered also as a generating function. Using (\ref{nsym1b})), we
express the generating functions in terms of
\begin{eqnarray}
\ (\Psi ^{2}){^{\bullet }} &=&-\int dt\ \Upsilon \ h{_{3}^{\bullet }}=-\int
dt\ (~^{F}\mathbf{\Upsilon }+~^{m}\mathbf{\Upsilon }+~\overline{\mathbf{%
\Upsilon }})\ h{_{3}^{\bullet }}  \label{genf2} \\
\Phi ^{2} &=&-4\Lambda h_{3}=-4(\ ^{F}\Lambda +\ ^{m}\Lambda +\ \overline{%
\Lambda })h_{3}.  \notag
\end{eqnarray}%
All formulas (\ref{offdcosm}), (\ref{nconnect}) and (\ref{genf2}) can be
rewirtten in terms of $\eta $-polarization functions d-metrics (\ref{dme}).
This allows us to compute in explicit form the d-metric and N-connection
coefficients, and respective nonlinear symmetries, for nonholonomic
deformations of certain given prime d-metrics into new classes of generic
off-diagonal, or diagonalized, cosmological solutions with "memory" on such
deformations.

We can chose some generating data $(\Psi ,\Upsilon ),$ or $(\Phi ,\Lambda ),$
related by nonlinear differential/integral symmetries (\ref{nsym1b}), and
respective integration functions in explicit form following certain
topology/ symmetry / asymptotic conditions for some classes of exact, or
small parametric, cosmological solutions. The coefficients (\ref{offdcosm})
define generic off-diagonal locally anisotropic cosmological solutions if
the corresponding anholonomy coefficients $C_{\alpha \beta }^{\gamma
}(x^{i},t)$ (see footnote \ref{ftnnonhcoef}) are not trivial. Such solutions
are with a nontrivial d-torsion induced nonholonomically N-adapted
coefficients which can be computed in explicit form. In order to generate as
particular cases some well-known cosmological FLRW, or Bianchi, type
metrics, we have to consider data of type $(\Psi (t),\Upsilon (t)),$ or $%
(\Phi (t),\Lambda ),$ with integration functions which allow frame/
coordinate transforms to respective (off-) diagonal configurations $%
g_{\alpha \beta }(t).$

\subsubsection{Extracting Levi-Civita STQC configurations}

\label{sssectlc}Such configurations with $\widehat{\mathbf{D}}_{\mid
\widehat{\mathcal{T}}\rightarrow 0}=\mathbf{\nabla }$ can be defined by
additional nonholonomic constraints (additional systems of PDEs on the
coefficients of the canonical d-metric and N-connection) or certain limits,
resulting in zero torsion $\widehat{\mathcal{T}}=0.$ Such LC--conditions for
(\ref{offdcosm}) are satisfied if
\begin{eqnarray}
\partial _{t}w_{i} &=&(\partial _{i}-w_{i}\partial _{t})\ln \sqrt{|h_{4}|}%
,(\partial _{i}-w_{i}\partial _{t})\ln \sqrt{|h_{3}|}=0,  \label{lca} \\
\partial _{k}w_{i} &=&\partial _{i}w_{k},\partial _{t}n_{i}=0,\partial
_{i}n_{k}=\partial _{k}n_{i}.  \notag
\end{eqnarray}%
This system of nonlinear first order PDEs can be solved in explicit form by
imposing additional nonholonomic constraints on cosmological d--metrics and
N-coefficients of (modified) Einstein equations. We can construct solutions
of (\ref{lca}) in explicit form if the generation functions and generating
sources in (\ref{offdcosm}) and (\ref{nconnect}) are subjected to certain
integrability conditions, for instance,
\begin{equation}
\Psi =\check{\Psi}(x^{i},t),(\partial _{i}\check{\Psi}){^{\bullet }}%
=\partial _{i}(\check{\Psi}{^{\bullet }})\mbox{ and }\Upsilon
(x^{i},t)=\Upsilon \lbrack \check{\Psi}]=\check{\Upsilon},\mbox{or }\Upsilon
=const,  \label{lc1}
\end{equation}%
where $\Upsilon \lbrack \check{\Psi}]$ sates that possible v-sources depends
via space-time coordinates only as functionals of $\check{\Psi}.$

Expressing
\begin{equation}
\check{h}_{3}=h_{3}^{[0]}(x^{k})-\int dt\frac{(\check{\Psi}^{2}){^{\bullet }}%
}{4\check{\Upsilon}}=h_{3}^{[0]}(x^{k})-\frac{\check{\Phi}^{2}}{4\Lambda },
\label{lcgf}
\end{equation}%
we can check by explicit computations that $\widehat{\mathcal{T}}=0$ if the
N-connection coefficients are taken

\begin{equation}
\begin{array}{c}
n_{k}=\check{n}_{k}=\partial _{k}n(x^{i})\mbox{ and } \\
\\
w_{i}=\check{w}_{i}=\partial _{i}\check{A}=\left\{
\begin{array}{c}
\frac{\partial _{i}(\int dt\ \check{\Upsilon}\ \check{h}_{3}^{\bullet })}{%
\check{\Upsilon}\ \check{h}_{3}^{\bullet }}; \\
\frac{\partial _{i}\check{\Psi}}{\check{\Psi}{^{\bullet }}}; \\
\frac{\partial _{i}[\int dt\ \check{\Upsilon}(\check{\Phi}^{2}){^{\bullet }}]%
}{\check{\Upsilon}(\check{\Phi}^{2}){^{\bullet }}}.%
\end{array}%
\right.%
\end{array}
\label{lc2}
\end{equation}%
A function $\check{A}(x^{i},t)$ can be prescribed in a necessary form which
restricts the class of possible generating functions subjected, for
instance, to the condition that it is a solution of the equation
\begin{equation}
(\partial _{i}\check{A})\check{\Psi}{^{\bullet }}=\partial _{i}\check{\Psi}.
\label{ztorsc}
\end{equation}%
Such solutions can be found in explicit or parametric forms if (inversely)
we prescribe $\check{\Psi}$ but compute $\check{A}$ as a solution of the
same equation.

For LC-configurations, d-metrics of type (\ref{dme1a}), (\ref{targ3a}) and (%
\ref{targ4b}) can be written respectively
\begin{eqnarray*}
ds^{2} &=&g_{i}(x^{k})[dx^{i}]^{2}+h_{3}(x^{k},t)[dy^{3}+\partial
_{k}n(x^{i})dx^{k}]^{2}+h_{4}(x^{k},t)[dt+\partial _{i}\check{A}%
(x^{k},t)dx^{i}]^{2} \\
&=&\eta _{i}\mathring{g}_{i}[\overline{e}^{i}]^{2}+\eta _{3}\mathring{h}_{3}
[\overline{\mathbf{e}}^{3}]^{2}+\mathring{h}_{4}[\overline{\mathbf{\check{e}}%
}^{4}]^{2} \\
&=&a^{2}(x^{k},\tau )\{\check{\eta}_{i}(x^{k},\tau )\mathring{g}_{i}[\check{e%
}^{i}]^{2}+\mathring{h}_{3}[\mathbf{\check{e}}^{3}]^{2}\}+\mathring{h}_{4}(%
\mathbf{\check{e}}^{4}(x^{k},\tau ))^{2}, \\
\mbox{ where } && \overline{e}^{i} = dx^{i},\overline{\mathbf{e}}%
^{3}=dy^{3}+\partial _{k}n(x^{i})dx^{k},\overline{\mathbf{\check{e}}}^{4}=%
\sqrt{|\eta _{4}|}(dt+\partial _{i}\check{A}(x^{k},t)dx^{k}), \\
&& \check{e}^{i} = dx^{i},\mathbf{\check{e}}^{3}=dy^{3}+\partial
_{k}n(x^{i})dx^{k},\mathbf{\check{e}}^{4}(x^{k},\tau )=d\tau +\partial _{i}%
\check{A}(x^{k},\tau )dx^{k}.
\end{eqnarray*}

For simplicity, in this work we shall provide formulas for canonical
d--connections considering that LC-configurations can be always extracted by
imposing additional constraints on generating/intergation functions and
sources as in formulas (\ref{lc1}), (\ref{lc2}) and (\ref{ztorsc}).

\subsubsection{Constructing d-metrics and N-connections from STQC generating
functions}

Cosmological d-metrics $\overline{\mathbf{g}}_{\alpha \beta }$ (\ref{stqcp})
describing STQC configurations can be generated by nonlinear off-diagonal
interactions of gravitational and matter fields. For our further
considerations, it is convenient to work with $\eta $--polarization
functions introduced for the d-metric (\ref{dme}). We shall overline such
values and generated coefficients of d-metrics and N-connections in order to
emphasize that they are generated for polarization functions constructed as
functionals on solutions for STQC structures. For instance, we shall write
\begin{equation}
\eta _{3}\rightarrow \overline{\eta }=\overline{\eta }_{3}({x}^{i},t)=%
\overline{\eta }_{3}[\overline{\varsigma },\overline{\chi },\overline{\breve{%
q}},\overline{\psi },\overline{q},...]  \label{gendeffunct}
\end{equation}%
depending on certain data $\overline{\varsigma }(x^{i},y^{a})$ (\ref{1tqc}),
$\overline{\chi }(x^{i},y^{a})$ and $\overline{\breve{q}}(x^{i},y^{a})$ from
(\ref{2tqc}), $\overline{\psi }(x^{i},y^{a})$ (\ref{qcevoleq}); and/or $%
\overline{\chi }(x^{i},y^{a})$ and $\overline{q}(x^{i},y^{a})$ from see (\ref%
{qevoleq}), defined by solutions of respective nonlinear dynamical and/or
evolution equations. The formula (\ref{coeffic}) for STQC nonholonomic
deformations in (\ref{dme}) is written in "overlined" form, $\overline{h}%
_{3}=\overline{a}^{2}=\overline{\eta }\mathring{a}^{2},$ where $\mathring{a}$
is a prime cosmological scaling factor taken, for instance, for a FLRW model.

The formulas for nonlinear symmetries of generating functions of STQC
systems (\ref{genf2}) can be written in terms of polarization function $%
\overline{\eta }_{4},$ generating (effective) matter sources, and
respectively associated (effective) cosmological constants,%
\begin{eqnarray*}
\ (\overline{\Psi }^{2}){^{\bullet }} &=&-\int dt\ \Upsilon \ (\overline{%
\eta }_{3}h{_{3})^{\bullet }}=-\int dt\ (~^{F}\mathbf{\Upsilon }+~^{m}%
\mathbf{\Upsilon }+~\overline{\mathbf{\Upsilon }})\ (\overline{\eta }_{3}h{%
_{3})^{\bullet }} \\
\overline{\Phi }^{2} &=&-4\Lambda (\overline{\eta }_{3}h{_{3})}=-4(\
^{F}\Lambda +\ ^{m}\Lambda +\ \overline{\Lambda })(\overline{\eta }_{3}h{%
_{3})}.
\end{eqnarray*}%
Such formulas allow us to rewrite, respectively, (\ref{offdcosm}) and (\ref%
{nconnect}) in terms of $\overline{\eta }$-polarizations,
\begin{eqnarray}
\overline{\eta }_{i}(x^{k},t) &=&\frac{g_{1}(x^{k})}{\mathring{g}%
_{i}(x^{k},t)}\mbox{ with }g_{1}=g_{2}(x^{k})%
\mbox{ as a solution of 2-d Poisson eqs.
with source }2~\Upsilon ;  \notag \\
\overline{\eta }_{3}(x^{k},t) &=&\overline{\eta }%
\mbox{ can be considered as
a  generating function, or }  \notag \\
\overline{\eta } &=&\mathring{h}_{3}^{-1}\left( h_{3}^{[0]}-\int dt\frac{(%
\overline{\Psi }^{2}){^{\bullet }}}{4\Upsilon }\right) =\mathring{h}%
_{3}^{-1}\left( h_{3}^{[0]}-\frac{\overline{\Phi }^{2}}{4\Lambda }\right) ;
\notag \\
\overline{\eta }_{4}(x^{k},t) &=&\frac{4[(|\overline{\eta }\mathring{h}%
_{3}|^{1/2}){^{\bullet }}]^{2}}{\mathring{h}_{4}|\int dt\Upsilon (\overline{%
\eta }h{_{3})^{\bullet }}|};  \notag \\
\overline{\eta }_{k}^{3}(x^{k},t) &=&\frac{_{1}n_{k}}{\mathring{n}_{k}}+4%
\frac{\ _{2}n_{k}}{\mathring{n}_{k}}\int dt\frac{[(|\overline{\eta }%
\mathring{h}_{3}|^{-1/4}){^{\bullet }}]^{2}}{|\int dt\Upsilon (\overline{%
\eta }h{_{3})^{\bullet }}|};\overline{\eta }_{i}^{4}(x^{k},t)=\frac{\partial
_{i}[\int dt\Upsilon (\overline{\eta }h{_{3})^{\bullet }}]}{\mathring{w}%
_{i}\Upsilon (\overline{\eta }h{_{3})^{\bullet }}}.  \label{stqcpolar}
\end{eqnarray}%
In these formulas, there are considered integration functions $%
_{1}n_{k}(x^{k}),$ $_{2}n_{k}(x^{k}),$ and $h_{4}^{[0]}(x^{k});$ and data
for a primary cosmological metric $\mathbf{\mathring{g}}$ (\ref{primedm}),
which can be prescribed, for instance, as a FLRW, or Bianchi type, d-metric.

Using formulas (\ref{coeffic}) and $\overline{\eta }$-polarizations (\ref%
{stqcpolar}), we can generate various classes of exact solutions of modified
Einstein equations (\ref{mfeq}). Such d-metrics and related N-connections
are parameterized in the form (\ref{dme}),
\begin{eqnarray}
\ \mathbf{\mathring{g}}\mathbf{\rightarrow }\overline{\mathbf{g}} &=&%
\overline{g}_{i}(x^{k})dx^{i}\otimes dx^{i}+\overline{h}_{a}(x^{k},t)%
\overline{\mathbf{e}}^{a}\otimes \overline{\mathbf{e}}^{a} = \overline{\eta }%
_{i}\mathring{g}_{i}dx^{i}\otimes dx^{i}+\overline{\eta }_{a}\mathring{h}_{a}%
\overline{\mathbf{e}}^{a}[\overline{\eta }]\otimes \overline{\mathbf{e}}^{a}[%
\overline{\eta }],  \notag \\
&&\mbox{ where }\overline{N}_{i}^{a}=\overline{\eta }_{i}^{a}\mathring{N}%
_{i}^{a}\mbox{ define } \overline{\mathbf{e}}^{a}[\overline{\eta }]=(dx^{i},%
\mathbf{e}^{a}=dy^{a}+\overline{\eta }_{i}^{a}\mathring{N}_{i}^{a}dx^{i}).
\label{stqcdme}
\end{eqnarray}%
The STQC character of such solutions is determined both by a gravitational
generating function taken as a polarization $\overline{\eta }$; and a
respective locally anisotropic cosmological scaling factor $\overline{a}^{2}=%
\overline{\eta }\mathring{a}^{2}$ (\ref{coeffic}) and by the components of a
generating source for (effective) matter $\Upsilon =$ $~^{F}\Upsilon
+~^{m}\Upsilon +~\overline{\mathbf{\Upsilon }}.$ We emphasize that
gravitational STQC configurations can be generated by usual matter fields, $%
~^{m}\Upsilon ,$ and effective sources in MGTs, $~^{F}\Upsilon ,$ even the
STQC matter source is taken$~\overline{\mathbf{\Upsilon }}=0.$

We also note that generic off-diagonal nonlinear interactions may generate
STQC configurations in GR for $~^{F}\Upsilon =0.$ The LC-configurations are
extracted as in previous subsection \ref{sssectlc} when, for instance, we
have to solve the equation (\ref{ztorsc}) rewritten for STQC generating
functions, $(\partial _{i}\overline{\check{A}})\overline{\check{\Psi}}{%
^{\bullet }}=\partial _{i}\overline{\check{\Psi}}.$ This restricts the class
of possible generating polarization functions $\overline{\eta }%
_{a}\rightarrow \overline{\check{\eta}}_{a}$ and generating sources $\check{%
\Upsilon}.$ In general, cosmological STQC-configurations are described by
generic off-diagonal metrics. Nevertheless, we can always prescribe/extract
diagonal configurations as it is described by formulas for small $%
\varepsilon $-deformations\ (\ref{targdm4c})--(\ref{targdm4f}). Quasi FLRW
configurations can be generated also by a subclass of generating functions
and sources $\overline{\check{\Psi}}(t)$ and $\check{\Upsilon}(t).$ If we
consider such prescriptions before constructing a general off-diagonal
locally anisotropic cosmological solution, we restrict our considerations
only to trivial nonholonomic deformations of diagonal cosmological metrics
into another cosmological metrics $g_{\alpha }(t).$ In such cases, we lose
the possibility to find cosmological solutions of (modified) Einstein
equations describing certain STQC or QC structures. The key point is to
apply the AFDM as we explained in details in \cite%
{cosmv2,eegrg,cosmv3,vbubuianu17} and construct general off-diagonal
d-metrics. At the end, we can consider $\varepsilon $-deformations and/or
constraints to $\overline{\check{\Psi}}(t)$ and $\check{\Upsilon}(t),$ with $%
h_{t}^{[0]}=const$ and $_{1}n_{k}(x^{k})=0$ and $_{2}n_{k}(x^{k})=0,$ which
would result in (diagonal) quasi FLRW metrics. This encodes STQC structures
(with a memory/code for more general off-diagonal nonlinear interactions)
via constants and parameters introduced for STQC dynamical and/or evolution
equations, for instance, of type (\ref{1tqc}), (\ref{2tqc}), (\ref{qcevoleq}%
); and/or (\ref{qevoleq}).

\subsubsection{Off-diagonal quadratic line elements with nonholonomic
STQC-torsion}

Putting together in (\ref{stqcdme}) the STCS $\overline{\eta }$-polarization
functions (\ref{stqcpolar}), we obtain general formulas for quadratic line
elements determined by off--diagonal cosmological solutions with Killing
symmetry on $\partial _{3}$ determined by sources, $~\ _{h}\Upsilon
=~_{h}^{F}\Upsilon +~_{h}^{m}\Upsilon +~\ _{h}\overline{\mathbf{\Upsilon }}$
and $\Upsilon =~^{F}\Upsilon +~^{m}\Upsilon +~\overline{\mathbf{\Upsilon }}$%
, and an effective nontrivial cosmological constant $\Lambda =\ ^{F}\Lambda
+\ ^{m}\Lambda +\ \overline{\Lambda }.$ In explicit form,
\begin{eqnarray}
ds^{2} &=&g_{1}(x^{k})[\overline{e}^{i}]^{2}-\overline{\eta }\mathring{h}%
_{3}[\overline{\mathbf{e}}^{3}]^{2}+\mathring{h}_{4}[\overline{\mathbf{e}}%
^{4}]^{2}  \label{qlestqc} \\
&=&\check{a}^{2}(x^{k},\tau )\{\overline{\check{\eta}}_{i}(x^{k},\tau )%
\mathring{g}_{i}[\check{e}^{i}]^{2}+\mathring{h}_{3}[\overline{\mathbf{%
\check{e}}}^{3}]^{2}\}+\mathring{h}_{4}(\overline{\mathbf{\check{e}}}%
^{4})^{2},  \notag
\end{eqnarray}%
where
\begin{eqnarray}
\overline{e}^{i} &=&\check{e}^{i}=dx^{i},  \notag \\
\overline{\mathbf{e}}^{3} &=&dy^{3}+(\frac{_{1}n_{k}}{\mathring{n}_{k}}+4%
\frac{\ _{2}n_{k}}{\mathring{n}_{k}}\int dt\frac{[(|\overline{\eta }%
\mathring{h}_{3}|^{-1/4}){^{\bullet }}]^{2}}{|\int dt(~^{F}\Upsilon
+~^{m}\Upsilon +~\overline{\mathbf{\Upsilon }})(\overline{\eta }h{%
_{3})^{\bullet }}|})\mathring{N}_{k}^{3}dx^{k},  \notag \\
\overline{\mathbf{e}}^{4} &=&\sqrt{|\overline{\eta }_{4}|}[dt+\frac{\partial
_{i}[\int dt\Upsilon (\overline{\eta }h{_{3})^{\bullet }}]}{\mathring{w}%
_{i}\Upsilon (\overline{\eta }h{_{3})^{\bullet }}}\mathring{N}%
_{k}^{4}dx^{k}],  \notag \\
\overline{\mathbf{\check{e}}}^{3} &=&dy^{3}+\overline{\check{\eta}}%
_{k}^{3}(x^{k},\tau )\mathring{N}_{k}^{3}dx^{k},\overline{\mathbf{\check{e}}}%
^{4}=d\tau +\overline{\check{\eta}}_{k}^{4}(x^{k},\tau )\mathring{N}%
_{k}^{4}dx^{k}.  \notag
\end{eqnarray}

The coefficients for polarization functions in (\ref{qlestqc}) are defined
and computed following formulas (\ref{targ4}) but considering STCS $%
\overline{\eta }$-polarization functions (\ref{stqcpolar}), when
\begin{eqnarray*}
\overline{\eta }_{3}(x^{k},t(x^{i},\tau )) &=&\overline{\eta }%
(x^{k},t(x^{i},\tau )),\ \mbox{ see formulas
(\ref{gendeffunct})}; \\
\check{a}^{2}(x^{k},\tau ):= &&\overline{\eta }_{3}(x^{k},t(x^{i},\tau ))%
\mathring{a}^{2}(x^{k},t(x^{i},\tau ))=\overline{\eta }(x^{k},t(x^{i},\tau ))%
\mathring{a}^{2}(x^{k},t(x^{i},\tau ))=\overline{\eta }(x^{k},\tau )%
\mathring{a}^{2}(x^{i},\tau ); \\
\overline{\check{\eta}}_{i}(x^{k},\tau ):= &&\frac{\overline{\eta }%
_{i}(x^{k},t(x^{i},\tau ))}{\overline{\eta }(x^{k},t(x^{i},\tau ))};\
\overline{\check{\eta}}_{k}^{3}(x^{k},\tau ):=\overline{\eta }%
_{k}^{3}(x^{k},t(x^{i},\tau )); \\
\overline{\check{\eta}}_{k}^{4}(x^{k},\tau ):= &&\overline{\eta }%
_{4}\{\partial _{k}t(x^{i},\tau )[\mathring{N}_{k}^{4}(x^{i},t(x^{i},\tau
))]^{-1}+\overline{\eta }_{k}^{4}(x^{i},t(x^{i},\tau ))\}\mathring{N}%
_{k}^{4}(x^{i},t(x^{i},\tau )).
\end{eqnarray*}

Such locally anisotropic and inhomogeneous cosmological solutions are, in
general, with nontrivial nonholonomically induced torsion. This class of
solutions can be re-defined equivalently in terms of generating functions $%
\overline{\Psi }({x}^{k},t)$ and/or $\overline{\Phi }({x}^{k},t),$ see $%
\overline{\eta }_{4}(x^{k},t)=\overline{\eta }_{4}[\overline{\Psi }]=%
\overline{\eta }_{4}[\overline{\Phi }]$ in (\ref{gendeffunct}).

\subsubsection{Quadratic elements for quasi FLRW solutions encoding STQC
structures}

Choosing integration functions resulting in zero N-connection coefficients,
we transform (\ref{qlestqc}) into a quasi FLRW metric of type (\ref{qflrwm}%
),
\begin{equation}
ds^{2}=\overline{\eta }(x^{k},\tau )\mathring{a}^{2}(x^{k},\tau )\{\overline{%
\check{\eta}}_{i}(x^{k},\tau )\mathring{g}_{i}[dx^{i}]^{2}+\mathring{h}%
_{3}[dy^{3}]^{2}\}+\mathring{h}_{4}[d\tau ]^{2}.  \label{qflrewstqc}
\end{equation}%
In such a d-metric, the horizontal polarizations $\overline{\check{\eta}}%
_{i}\ $are determined by a 2-d Poisson equation with source $~\ _{h}\Upsilon
$ \ as it was explained in (\ref{offdcosm}). The modifications of the
cosmological factor $\check{a}^{2}(x^{k},\tau )=\overline{\eta }(x^{k},\tau )%
\mathring{a}^{2}(x^{k},\tau )$ are given by the polarization $\overline{\eta
}(x^{k},\tau )$ in (\ref{stqcpolar}) determined by solutions of a nonlinear
system with source $\Upsilon (x^{k},\tau )$ and STQC generating functions $%
\overline{\Psi }({x}^{k},t)$ and/or $\overline{\Phi }({x}^{k},t).$

We can consider subclasses of generating functions and sources depending
only on time like coordinate, when and construct subclasses of solutions of
type%
\begin{equation}
ds^{2}=\overline{\eta }(\tau )\mathring{a}^{2}(\tau )\{\overline{\check{\eta}%
}_{i}(\tau )\mathring{g}_{i}[dx^{i}]^{2}+\mathring{h}_{3}(dy^{3})^{2}\}+%
\mathring{h}_{4}(d\tau )^{2}.  \label{qflrewtqc}
\end{equation}%
Such diagonal metrics describe cosmological spacetimes with TC strucure or
with memory of STQC, TQC and other type nonlinear interactions. We can chose
nonlinear coordinates when $\overline{\check{\eta}}_{i}(\tau )\simeq
1+\varepsilon \check{\chi}_{i}(\tau )$ as in (\ref{targdm4f}).

Using effective scale factors $\check{a}^{2}(x^{k},\tau )$ and/or $\check{a}%
^{2}(\tau ),$ respectively, in (\ref{qflrewstqc}) and/or (\ref{qflrewtqc}),
we can introduce the respective effective and locally anisotropically
polarized Hubble parameters,
\begin{eqnarray}
\check{H}:= &\partial (\ln \check{a})/\partial \tau =&(\ln \check{a}){%
^{\bullet }=}(\ln \mathring{a}){^{\bullet }+}(\ln \sqrt{|\overline{\eta }|}){%
^{\bullet }},  \label{hablcdef} \\
&=&\mathring{H}+~^{\eta }H,\mbox{ for }\mathring{H}:=(\ln \mathring{a}){%
^{\bullet }\mbox{ and }}~^{\eta }H:=(\ln \sqrt{|\overline{\eta }|}){%
^{\bullet }.}  \notag
\end{eqnarray}%
Such a $\check{H}(x^{k},\tau ),$ or $\check{H}(\tau ),\ $ encodes
information on gravitational and matter field STQC structures determined by
corresponding classes of dynamical and evolution equations. Cosmological
(off-) diagonal evolution in MGTs and GR is defined by polarization $%
\overline{\eta }=\overline{\eta }_{3}$ computed by formula%
\begin{eqnarray}
\overline{\eta }(x^{k},\tau ) &=&h_{3}^{[0]}(x^{k})-\frac{\Phi
^{2}(x^{k},\tau )}{4\Lambda }\mbox{ or }  \label{polarizf} \\
\overline{\eta }(\tau ) &=&h_{3}^{[0]}-\frac{\Phi ^{2}(\tau )}{4\Lambda }%
,h_{3}^{[0]}=const,\mbox{ for deformations of flat FLRW
metrics, }\mathring{h}_{4}=-1,  \notag
\end{eqnarray}%
Other polarizations describe, in general, a locally anisotropic and/or
inhomogeneous gravitational background. In these formulas, the generating $%
\Upsilon $-sources are prescribed for (effective) matter and STQC fields and
the polarization/function $\overline{\eta }_{4}$ determine nonlinear
gravitational interactions as solutions of (modified) Einstein equations.

We can compute nonholonomic deformations of the prime Hubble radius $%
\mathring{R}_{H}$ to a target effective Hubble radius $\check{R}_{H},$
\begin{equation*}
\mathring{R}_{H}:=\frac{1}{\mathring{a}\mathring{H}}\rightarrow \check{R}%
_{H}:=\frac{1}{\check{a}\check{H}}=\mathring{R}_{H}\frac{1}{\sqrt{|\overline{%
\eta }|}+(\sqrt{|\overline{\eta }|}){^{\bullet }/}\mathring{H}}.
\end{equation*}%
The polarization $\overline{\eta }$ modifies the conditions ($\check{R}_{H}{%
^{\bullet }}<0$ implying $\check{a}{^{\bullet \bullet }>0)}$ that have to be
satisfied in order for inflation (also called early-time accelerating era)
to occur in a STQC media.

\subsection{Examples of STQC, time QC and TC solutions}

\label{ssexamplsol}In this subsection, we compute in explicit form certain
examples of conformal factor polarizations generating STQC structures.

\subsubsection{Gravitational STQC structures generated by (effective) matter
fields}

Any matter type source $\Upsilon $ modifies nonholonomically a prime
cosmological factor $\mathring{a}$ and could generate quasiperiodic
space-time structures for certain well-defined conditions. Let us analyze
some classes of such solutions:

\begin{enumerate}
\item \textbf{STQC from standard matter in GR:}\newline
A source$~^{m}\mathbf{\Upsilon }_{\mu \nu }=(2M_{P}^{2})^{-1}\ ^{m}\mathbf{T}%
_{\mu \nu }$ (\ref{msourc}) [we can consider, for instance, $\ ^{m}\mathbf{T}%
_{\mu \nu }$ for a scalar field] defines a STQC polarization of type (\ref%
{polarizf}),%
\begin{equation*}
\overline{\eta }(x^{k},\tau )=\overline{\eta }=\mathring{h}%
_{3}^{-1}(h_{3}^{[0]}-\int dt\frac{(\overline{\Psi }^{2}){^{\bullet }}}{%
4~^{m}\Upsilon })=\mathring{h}_{3}^{-1}(h_{3}^{[0]}-\frac{\overline{\Phi }%
^{2}}{4\ ^{m}\Lambda }),
\end{equation*}%
if the generating functions (\ref{lcgf}) are solutions of corresponding STQC
equations. We can model such configurations for different choices of
functionals for generating functions and $\mathring{h}_{4}=-1,$
\begin{equation*}
\overline{\eta }_{3}=h_{3}^{[0]}(x^{k})-\int d\tau \frac{(\check{\Psi}^{2}){%
^{\bullet }}}{4~^{m}\check{\Upsilon}},\mbox{ where }\left\{
\begin{array}{ccc}
\check{\Psi}=\check{\Psi}[\overline{\varsigma }], & \mbox{ 1TC, } &
\overline{\varsigma }(x^{i},\tau )\mbox{ from (\ref{1tqc})}; \\
\check{\Psi}=\check{\Psi}[\overline{\chi },\overline{\breve{q}}], & %
\mbox{2TC,} & \overline{\chi }(x^{i},\tau ),\overline{\breve{q}}(x^{i},\tau )%
\mbox{ from (\ref{2tqc})}; \\
\check{\Psi}=\check{\Psi}[\overline{\psi }], & \mbox{QC,} & \overline{\psi }%
(x^{i},\tau )\mbox{ from (\ref{qcevoleq}) }; \\
\check{\Psi}=\check{\Psi}[\overline{\chi},\overline{q}], & \mbox{mix QC,TCQ,}
& \overline{\chi}(x^{i},\tau),\overline{q}(x^{i},\tau)\mbox{from(%
\ref{qevoleq})}.%
\end{array}%
\right.
\end{equation*}%
We model a cosmological evolution only with time dependent d-metrics with
memory on STQC structure of nonlinear interactions if we consider generation
functions, or limits to configurations, for
\begin{equation*}
\overline{\eta }_{3}=h_{3}^{[0]}-\int d\tau \frac{(\check{\Psi}^{2}(\tau )){%
^{\bullet }}}{4~^{m}\check{\Upsilon}(\tau )},\mbox{ where }\left\{
\begin{array}{ccc}
\check{\Psi}=\check{\Psi}[\overline{\varsigma }], & \mbox{ 1TC, } &
\overline{\varsigma }(\tau ); \\
\check{\Psi}=\check{\Psi}[\overline{\chi },\overline{\breve{q}}], & %
\mbox{2TC,} & \overline{\chi }(\tau ),\overline{\breve{q}}(\tau ); \\
\check{\Psi}=\check{\Psi}[\overline{\psi }], & \mbox{QC,} & \overline{\psi }%
(\tau ); \\
\check{\Psi}=\check{\Psi}[\overline{\chi },\overline{q}], &
\mbox{mix
QC,TCQ,} & \overline{\chi }(\tau ),\overline{q}(\tau ).%
\end{array}%
\right.
\end{equation*}%
The integration constants can be prescribed in certain forms when there are
generated observational quasiperiodic and pattern-forming structures.

\item \textbf{STQC from effective sources in MGTs:} \newline
A $F(R)$ modification of GR induces an effective matter source$\ \ ^{F}%
\mathbf{\Upsilon }_{\mu \nu }$ (\ref{fsourc}) which results in STQC
polarizations of type (\ref{polarizf}),%
\begin{equation*}
\overline{\eta }(x^{k},\tau )=\overline{\eta }=\mathring{h}%
_{3}^{-1}(h_{3}^{[0]}-\int dt\frac{(\overline{\Psi }^{2}){^{\bullet }}}{%
4~^{F}\Upsilon })=\mathring{h}_{3}^{-1}(h_{3}^{[0]}-\frac{\overline{\Phi }%
^{2}}{4\ ^{F}\Lambda }),
\end{equation*}
if the generating functions in (\ref{offdcosm}) are solutions of
corresponding STQC equations. In explicit form, such modifying gravity
generating functions can be computed for $\mathring{h}_{4}=-1,$
\begin{equation}
\overline{\eta }_{3}=h_{3}^{[0]}(x^{k})-\int d\tau \frac{(\overline{\Psi }%
^{2}){^{\bullet }}}{4~\ ^{F}\mathbf{\Upsilon }},\mbox{ where }\left\{
\begin{array}{ccc}
\overline{\Psi }=\Psi \lbrack \overline{\varsigma }], & \mbox{ 1TC, } &
\overline{\varsigma }(x^{i},\tau ); \\
\overline{\Psi }=\Psi \lbrack \overline{\chi },\overline{\breve{q}}], & %
\mbox{2TC,} & \overline{\chi }(x^{i},\tau ),\overline{\breve{q}}(x^{i},\tau
); \\
\overline{\Psi }=\Psi \lbrack \overline{\psi }], & \mbox{QC,} & \overline{%
\psi }(x^{i},\tau ); \\
\overline{\Psi }=\Psi \lbrack \overline{\chi },\overline{q}], &
\mbox{mix
QC,TCQ,} & \overline{\chi }(x^{i},\tau ),\overline{q}(x^{i},\tau ).%
\end{array}%
\right.  \label{genfstqc}
\end{equation}%
Such configurations are, in general, with nontrivial canonical torsion $%
\widehat{\mathcal{T}}=0.$ Chosing generating functions $\check{\Psi}$ and
considering limits $\widehat{\mathbf{D}}_{\mid \widehat{\mathcal{T}}=0}=%
\mathbf{\nabla ,}$ we can select LC-configurations for cosmological
solutions in MGT. \ Time $\tau $-depending metrics can be generated for
configurations with $\overline{\varsigma }(\tau );\overline{\chi }(\tau ),%
\overline{\breve{q}}(\tau );\overline{\psi }(\tau );$ or $\overline{\chi }%
(\tau ),\overline{q}(\tau ).$

\item \textbf{Gravitational} \textbf{STQC configurations from quasiperiodic
matter :} \newline
A d-metric $\overline{\mathbf{g}}_{\alpha \beta }[x^{i},y^{a};\overline{%
\varsigma },\overline{\chi },\overline{\breve{q}},\overline{\psi },\overline{%
q},...]$ (\ref{stqcp}) can be generated by a source $\overline{\mathbf{%
\Upsilon }}_{\mu \nu }=\frac{1}{2M_{P}^{2}}\ \overline{\mathbf{T}}_{\mu \nu
} $ $\ $(\ref{msourc}) even generating functions are prescribed not in a
STQC form. \ Such polarizations functions (\ref{polarizf}) are determined by
$\eta _{3}(x^{i},y^{a})$ and/or $\Psi (x^{i},y^{a}),$ respectively, $\eta
_{3}(\tau )$ and/or $\Psi (\tau ),$
\begin{eqnarray}
\overline{\eta }_{3}(x^{k},\tau ) &=&h_{3}^{[0]}(x^{k})-\int d\tau \frac{%
\lbrack \Psi ^{2}(x^{i},\tau )]{^{\bullet }}}{4~\ ~\overline{\mathbf{%
\Upsilon }}[x^{i},\tau ;\overline{\varsigma },\overline{\chi },\overline{%
\breve{q}},\overline{\psi },\overline{q}]},\mbox{ or }  \notag \\
\overline{\eta }_{3}(\tau ) &=&h_{3}^{[0]}-\int d\tau \frac{\lbrack \Psi
^{2}(\tau )]{^{\bullet }}}{4~\ ~\overline{\mathbf{\Upsilon }}[\tau ;%
\overline{\varsigma }(\tau ),\overline{\chi }(\tau ),\overline{\breve{q}}%
(\tau ),\overline{\psi }(\tau ),\overline{q}(\tau )]}  \label{stqccosm} \\
&&\mbox{ for deformations of flat FLRW space-times with }\mathring{h}_{4}=-1.
\notag
\end{eqnarray}%
LC-conditions on this class of d-metric should be imposed and solved
additionally.
\end{enumerate}

\subsubsection{Interacting STQC gravitational and matter fields}

Let us analyze two classes of of nonholonomic deformations:

\begin{itemize}
\item In formulas (\ref{stqccosm}), we can introduce STQC generating
functions (\ref{genfstqc}) instead of $\Psi (x^{i},\tau )$ and compute
\begin{eqnarray}
\overline{\eta }_{3}(x^{k},\tau ) &=&h_{3}^{[0]}(x^{k})-\int d\tau \frac{%
\left( \overline{\Psi }^{2}[\overline{\varsigma }(x^{i},\tau ),\overline{%
\chi }(x^{i},\tau ),\overline{\breve{q}}(x^{i},\tau ),\overline{\psi }%
(x^{i},\tau ),\overline{q}(x^{i},\tau )]\right) {^{\bullet }}}{4~\ ~%
\overline{\mathbf{\Upsilon }}[\overline{\varsigma }(x^{i},\tau ),\overline{%
\chi }(x^{i},\tau ),\overline{\breve{q}}(x^{i},\tau ),\overline{\psi }%
(x^{i},\tau ),\overline{q}(x^{i},\tau )]},\mbox{ or }  \notag \\
\overline{\eta }_{3}(\tau ) &=&h_{3}^{[0]}-\int d\tau \frac{\lbrack
\overline{\Psi }^{2}[\overline{\varsigma }(\tau ),\overline{\chi }(\tau ),%
\overline{\breve{q}}(\tau ),\overline{\psi }(\tau ),\overline{q}(\tau )]]{%
^{\bullet }}}{4~\ ~\overline{\mathbf{\Upsilon }}[\overline{\varsigma }(\tau
),\overline{\chi }(\tau ),\overline{\breve{q}}(\tau ),\overline{\psi }(\tau
),\overline{q}(\tau )]}  \notag \\
&&\mbox{ for deformations of flat FLRW
space-times with  }\mathring{h}_{4}=-1.  \label{stqccosm1}
\end{eqnarray}

\item We can study STQC gravitational interactions with effective
cosmological constant $\Lambda =\ ^{F}\Lambda +\ ^{m}\Lambda +\ \overline{%
\Lambda }$ as in (\ref{offdcosm})%
\begin{eqnarray*}
\overline{\eta }_{3}(x^{k},\tau ) &=&\mathring{h}_{3}^{-1}\left( h_{3}^{[0]}-%
\frac{\overline{\Phi }^{2}[\overline{\varsigma }(x^{i},\tau ),\overline{\chi
}(x^{i},\tau ),\overline{\breve{q}}(x^{i},\tau ),\overline{\psi }(x^{i},\tau
),\overline{q}(x^{i},\tau )]}{4(\ ^{F}\Lambda +\ ^{m}\Lambda +\ \overline{%
\Lambda })}\right) \mbox{ or } \\
\overline{\eta }_{3}(\tau ) &=&\mathring{h}_{3}^{-1}\left( h_{4}^{[0]}-\frac{%
\overline{\Phi }^{2}[\overline{\varsigma }(\tau ),\overline{\chi }(\tau ),%
\overline{\breve{q}}(\tau ),\overline{\psi }(\tau ),\overline{q}(\tau )]}{%
4(\ ^{F}\Lambda +\ ^{m}\Lambda +\ \overline{\Lambda })}\right) .
\end{eqnarray*}%
LC-configurations in GR can be extracted from these formulas if we prescribe
$\ ^{F}\Lambda =0$ and $\overline{\Phi }[\overline{\varsigma },\overline{%
\chi },\overline{\breve{q}},\overline{\psi },\overline{q}]$ is chosen as in (%
\ref{lcgf}). For such solutions, we can re-define formulas with explicit
sources $~^{F}\Upsilon +~^{m}\Upsilon +~\overline{\mathbf{\Upsilon }}$ if a
corresponding $\overline{\Psi }$--generating function is introduced if there
are considered nonlinear symmetries (\ref{genf2}).
\end{itemize}

\subsubsection{A toy model with 1-TQC structure for DM}

\label{ssexamplsoltoy} A simple example of d-metric $\overline{\mathbf{g}}%
_{\alpha \beta }[x^{i},y^{a};\varsigma ,...]$ of type (\ref{stqcp}) can be
generated by a source $\overline{\mathbf{\Upsilon }}_{\mu \nu }[\varsigma ]=%
\frac{1}{2M_{P}^{2}}\ \overline{\mathbf{T}}_{\mu \nu }[\varsigma ],$ see
formulas (\ref{msourc}) and (\ref{efstemt}), computed for a DM Lagrange
density $\overline{\mathcal{L}}=\acute{L}(\varsigma ),$ see formulas (\ref%
{sumlagd}) and (\ref{1tqclg}). In such a toy model, the generating functions
are prescribed in a general but not in a STQC form. Nevertheless, the
polarizations functions (\ref{polarizf}) are determined by $\eta
_{3}(x^{i},y^{a})$ and/or $\Psi (x^{i},y^{a}),$ respectively, $\eta
_{3}(\tau )$ and/or $\Psi (\tau ),$ as a subclass of configurations of type (%
\ref{stqccosm}) when
\begin{eqnarray}
\overline{\eta }_{3}[x^{k},\tau ;\varsigma (x^{k},\tau )]
&=&h_{3}^{[0]}(x^{k})-\int d\tau \frac{\lbrack \Psi ^{2}(x^{i},\tau )]{%
^{\bullet }}}{4~\ ~\overline{\mathbf{\Upsilon }}[x^{i},\tau ;\varsigma ]},%
\mbox{ or }  \notag \\
\overline{\eta }_{3}[\tau ;\varsigma (\tau )] &=&h_{3}^{[0]}-\int d\tau
\frac{\lbrack \Psi ^{2}(\tau )]{^{\bullet }}}{4~\ ~\overline{\mathbf{%
\Upsilon }}[\tau ;\varsigma (\tau )]}  \label{toysol} \\
&&\mbox{ for deformations of flat FLRW space-times with }\mathring{h}_{4}=-1.
\notag
\end{eqnarray}%
We can impose the LC-conditions on this class of d-metric and generate such
1-TQC configurations for DM in the framework of GR.

Off-diagonal and diagonalized cosmological metrics with gravitational
polarizations of type (\ref{toysol}) results in nontrivial inflation
scenarios as we conclude below in subsection \ref{ssinfls1tc}. Such values
of $\overline{\eta }_{3}[\varsigma ]$ can be used also for elaborating
examples of toy models of a late 1-TQC dynamics with some general DE but
just 1-TQC configurations for DM. We should reproduced for such a particular
choice all constructions from section \ref{s5} where the approach is
elaborated for a general STQC dynamics. The simplified scenarios of 1-TQC
acceleration and DM are useful in understanding possible physical
implications of cosmological solutions with time quaisperiodic structures
both in MGTs and GR. Nevertheless, modern cosmological data \cite%
{plank,plank1} state a more complex nonlinear structure and evolution of
Universe which can be described more realistically if there are used 2-TQC
and more sophisticate time and space like quasiperiodic configurations and
interactions of gravitational and (effective) matter fields.

\section{Inflationary STQC dynamics}

\label{s4} We studied models of in MGTs and GR for locally anisotropic
space-time quasi-periodic and pattern forming structures in \cite%
{cosmv2,eegrg,cosmv3,rajpvcosm,amaral17,vbubuianu17}, see references
therein. The $F(R)$ gravity is the most popular among MGTs and admits
various massive, mimetic, topological gravity and string type
generalizations - there are a series of reviews in literature \cite%
{capoz,nojod1,gorbunov,clifton,basilakos,linde2,gwyn,sami,mavromatos,nojiri,vagnozzi}%
. The goal of this section is to elaborate on inflation cosmology in
nonholonomic space-times with STQC structure encoded into an effective
cosmological scale factor $\overline{a}^{2}=\overline{\eta }\mathring{a}^{2}$
and corresponding nonholonomic deformations of Hubble constants, $\check{H}=%
\mathring{H}+~^{\eta }H$ (\ref{hablcdef}), with gravitational polarization $%
\overline{\eta }$ (\ref{polarizf}). We can always chose generating functions
or find limits to configurations with $\check{H}(\tau )=\mathring{H}(\tau
)+~^{\eta }H(\tau )$ and $\overline{\eta }(\tau )$ encoding prescribed STQC
structures as we described in previous section. These values can be used for
modeling inflation scenarios in off-diagonal backgrounds with nonholonomic
N-connection structures and/or "memory" of nonlinear gravitational and
matter field interactions contained into diagonalized quasi FLRW metrics (%
\ref{qflrwm}).

\subsection{General cosmological properties of gravity theories with STQC
structure}

\subsubsection{Perfect fluid representation of STQC configurations in GR}

We start with the quasi FLRW configurations in GR coupled with an effective
perfect fluid modelling STQC structures (with nontrivial effective sources $%
~^{m}\Upsilon $ and$~\overline{\mathbf{\Upsilon }}$ but $~^{F}\Upsilon =0,$
from for gravitational field equations (\ref{mfeq})). With respect to
N-adapted frames, we have (see (\ref{sourcc}))
\begin{equation*}
T_{\alpha \beta }\sim diag[\overline{\check{\eta}}_{i}(\tau
)~^{e}p,~^{e}p,-~^{e}p],
\end{equation*}%
for respective components of effective pressure, $~^{e}p=~^{m}p+~\overline{p}%
, $ and effective energy density, $~^{e}\rho =$ $~^{m}\rho +~\overline{\rho }%
,$ components. The polarizations $\overline{\check{\eta}}_{i}(\tau )\simeq
1+\varepsilon \overline{\chi }_{i}(\tau )$ can be chosen and treated as
small anisotropies for diagonalized d-metrics, see (\ref{targdm4f}). For
such parameterizations, the effective FLRW equations for diagonalized target
d-metrics are%
\begin{eqnarray}
~^{m}p+~\overline{p} &=&\frac{3}{2M_{P}^{2}}\check{H}^{2}=\frac{3}{2M_{P}^{2}%
}(\mathring{H}+~^{\eta }H)^{2},  \label{effriedman} \\
~^{m}\rho +~\overline{\rho } &=&-\frac{1}{2M_{P}^{2}}[3\check{H}^{2}+2\check{%
H}{^{\bullet }}]=-\frac{1}{2M_{P}^{2}}[3(\mathring{H}+~^{\eta }H)^{2}+2(%
\mathring{H}+~^{\eta }H){^{\bullet }}].  \notag
\end{eqnarray}%
In GR, it is also used the gravitational constant $\kappa ^{2}=2M_{P}^{2}.$
The effective equation of state (EoS) is computed
\begin{equation*}
~^{e}w=-1-2\check{H}{^{\bullet }/}3\check{H}^{2}=-1-2(\mathring{H}+~^{\eta
}H){^{\bullet }/}3(\mathring{H}+~^{\eta }H)^{2}.
\end{equation*}%
Such formulas encode STQC configurations even if $\ $quasi-periodic sources
of matter are with$~\overline{\rho }=0$ and $\overline{p}=0$ but nontrivial
values $~^{\eta }H$ are defined by gravitational STQC interactions

\subsubsection{Analogous perfect fluid representation of STQC structures in
MGT}

For flat quasi FLRW metrics (\ref{qlestqc}), the modified gravity equations (%
\ref{mfeq}) with sources (\ref{sourcc}) results in FLRW equations (see
details in \cite{cosmv2,eegrg,cosmv3,rajpvcosm,vbubuianu17}; for simplicity,
we use only the effective Hubble rate $\check{H}$ and $\mathbf{F}(\widehat{%
\mathbf{R}})$ and it functional derivatives on $\widehat{\mathbf{R}}\simeq 12%
\check{H}^{2}+6\check{H}{^{\bullet },}$ $\check{H}{^{\bullet }=\partial
\check{H}/\partial \tau ,}$ which can subjected to additional LC-conditions),%
\begin{eqnarray*}
2M_{P}^{2}(~^{m}p+~\overline{p}) &=&-36[\check{H}{^{\bullet \bullet }+}4%
\check{H}\check{H}{^{\bullet }}]^{2}\mathbf{F}^{\prime \prime \prime }(%
\widehat{\mathbf{R}})-6[\check{H}{^{\bullet \bullet \bullet }+6}\check{H}{%
^{\bullet \bullet }}\check{H}+4(\check{H}{^{\bullet }})^{2}+8\check{H}^{2}%
\check{H}{^{\bullet }}]\mathbf{F}^{\prime \prime }(\widehat{\mathbf{R}}) \\
&&+(3\check{H}^{2}+\check{H}{^{\bullet }})\mathbf{F}^{\prime }(\widehat{%
\mathbf{R}})-\frac{1}{2}\mathbf{F}(\widehat{\mathbf{R}}), \\
2M_{P}^{2}(~^{m}\rho +~\overline{\rho }) &=&18[\check{H}{^{\bullet \bullet }}%
\check{H}+4(\check{H}{^{\bullet }})^{2}]\mathbf{F}^{\prime \prime }(\widehat{%
\mathbf{R}})-3(\check{H}{^{\bullet }+}\check{H}^{2})\mathbf{F}^{\prime }(%
\widehat{\mathbf{R}})+\frac{1}{2}\mathbf{F}(\widehat{\mathbf{R}}).
\end{eqnarray*}%
These equations can be represented in effective form (\ref{effriedman}),
\begin{equation*}
~\check{p}=\frac{3}{2M_{P}^{2}}\check{H}^{2}\mbox{ and }\check{\rho}=-\frac{1%
}{2M_{P}^{2}}(3\check{H}^{2}+2\check{H}{^{\bullet })},
\end{equation*}%
when the respective values are computed for $\mathbf{F}(\widehat{\mathbf{R}}%
)=\widehat{\mathbf{R}}+\mathbf{\check{F}}(\widehat{\mathbf{R}}),$
\begin{eqnarray*}
2M_{P}^{2}~\check{p} &=&2M_{P}^{2}(~^{m}p+~\overline{p})+36[\check{H}{%
^{\bullet \bullet }+}4\check{H}\check{H}{^{\bullet }}]^{2}\mathbf{\check{F}}%
^{\prime \prime \prime }(\widehat{\mathbf{R}})+6[\check{H}{^{\bullet \bullet
\bullet }+6}\check{H}{^{\bullet \bullet }}\check{H}+4(\check{H}{^{\bullet }}%
)^{2} \\
&&+8\check{H}^{2}\check{H}{^{\bullet }}]\mathbf{\check{F}}^{\prime \prime }(%
\widehat{\mathbf{R}})-(3\check{H}^{2}+\check{H}{^{\bullet }})\mathbf{\check{F%
}}^{\prime }(\widehat{\mathbf{R}})+\frac{1}{2}\mathbf{\check{F}}(\widehat{%
\mathbf{R}}), \\
2M_{P}^{2}\check{\rho} &=&2M_{P}^{2}(~^{m}\rho +~\overline{\rho })-18[\check{%
H}{^{\bullet \bullet }}\check{H}+4(\check{H}{^{\bullet }})^{2}]\mathbf{%
\check{F}}^{\prime \prime }(\widehat{\mathbf{R}})+3(\check{H}{^{\bullet }+}%
\check{H}^{2})\mathbf{\check{F}}^{\prime }(\widehat{\mathbf{R}})-\frac{1}{2}%
\mathbf{\check{F}}(\widehat{\mathbf{R}}).
\end{eqnarray*}

For LC-configurations, these formulas admit a scalar tensor description and
can be used elaborating viable dark energy models, see review of works in
\cite{nojod1}. In \cite{cosmv2,cosmv3,vbubuianu17}, there are studied
off-diagonal generalizations for locally anisotropic cosmological models
with possible massive gravity and QC generalizations.

\subsubsection{Non-minimal coupling of MGT and STQC configurations}

We studied locally anisotropic cosmological configurations and/or QC
structures in MGTs with non-minimal coupling in \cite{eegrg,vbubuianu17}. In
this work, we elaborate on modified cosmology model with STQC structure
defined by action%
\begin{equation*}
~~^{f}\mathcal{S}=\int d^{4}u\sqrt{|\mathbf{g}|}\{\frac{M_{P}^{2}}{2}\mathbf{%
F}(\widehat{\mathbf{R}})+\mathbf{f}(\widehat{\mathbf{R}})(~^{m}\mathcal{L}+\
\overline{\mathcal{L}})\},
\end{equation*}%
where $\mathbf{f}(\widehat{\mathbf{R}})$ is a functional on $\widehat{%
\mathbf{R}}$ and Lagrange densities are used for definition of
energy-momentum tensors for usual matter and STQC matter sources (\ref%
{msourc}), $\ ^{m}\mathbf{T}_{\mu \nu }$ and $\ \overline{\mathbf{T}}_{\mu
\nu }$. Applying a N-adapted variational calculus for $\ ^{f}\mathcal{S},$
we obtain the gravitational field equations
\begin{equation}
\widehat{\mathbf{R}}_{\mu \nu }\mathbf{F}^{\prime }(\widehat{\mathbf{R}})+[%
\mathbf{g}_{\mu \nu }\widehat{\mathbf{D}}^{\tau }\widehat{\mathbf{D}}_{\tau
}-\frac{1}{2}(\widehat{\mathbf{D}}_{\mu }\widehat{\mathbf{D}}_{\nu }+%
\widehat{\mathbf{D}}_{\nu }\widehat{\mathbf{D}}_{\mu })])\mathbf{F}^{\prime
}(\widehat{\mathbf{R}})-\frac{1}{2}\mathbf{g}_{\mu \nu }\mathbf{F}(\widehat{%
\mathbf{R}})=\frac{1}{2M_{P}^{2}}\ ^{f}\mathbf{T}_{\mu \nu }  \label{nmceq}
\end{equation}%
where the effective energy-momentum tensor is defined
\begin{equation*}
\ ^{f}\mathbf{T}_{\mu \nu }:=[\ ^{m}\mathbf{T}_{\mu \nu }+\ \overline{%
\mathbf{T}}_{\mu \nu }-\widehat{\mathbf{R}}_{\mu \nu }(~^{m}\mathcal{L}+\
\overline{\mathcal{L}})]\mathbf{f}^{\prime }(\widehat{\mathbf{R}})+[\mathbf{g%
}_{\mu \nu }\widehat{\mathbf{D}}^{\tau }\widehat{\mathbf{D}}_{\tau }-\frac{1%
}{2}(\widehat{\mathbf{D}}_{\mu }\widehat{\mathbf{D}}_{\nu }+\widehat{\mathbf{%
D}}_{\nu }\widehat{\mathbf{D}}_{\mu })]\mathbf{f}^{\prime }(\widehat{\mathbf{%
R}})
\end{equation*}%
In the approximation of effective perfect fluid modelling matter field
interactions and STQC structures, we define the $f$-effective pressure and
energy density,
\begin{eqnarray*}
\ ^{f}p:= &&\frac{\partial ^{2}}{\partial \tau ^{2}}[\mathbf{f}^{\prime }(%
\widehat{\mathbf{R}})(~^{m}\mathcal{L}+\ \overline{\mathcal{L}})]+4\check{H}%
\frac{\partial }{\partial \tau }[\mathbf{f}^{\prime }(\widehat{\mathbf{R}}%
)(~^{m}\mathcal{L}+\ \overline{\mathcal{L}})] \\
&&-(\check{H}{^{\bullet }+3}\check{H}^{2})\mathbf{f}^{\prime }(\widehat{%
\mathbf{R}})(~^{m}\mathcal{L}+\ \overline{\mathcal{L}})+\mathbf{f}(\widehat{%
\mathbf{R}})(~^{m}p+~\overline{p}), \\
\ ^{f}\rho := &&-3\check{H}\frac{\partial }{\partial \tau }[\mathbf{f}%
^{\prime }(\widehat{\mathbf{R}})(~^{m}\mathcal{L}+\ \overline{\mathcal{L}}%
)]+3(\check{H}{^{\bullet }+}\check{H}^{2})\mathbf{f}^{\prime }(\widehat{%
\mathbf{R}})(~^{m}\mathcal{L}+\ \overline{\mathcal{L}})+\mathbf{f}(\widehat{%
\mathbf{R}})(~^{m}\rho +~\overline{\rho }),
\end{eqnarray*}%
where $(~^{m}p+~\overline{p})$ and $(~^{m}\rho +~\overline{\rho })$ are
defined by $\ ^{m}\mathbf{T}_{\mu \nu }+\overline{\mathbf{T}}_{\mu \nu }.$

It should be noted that models with non-trivial coupling of type $\ ^{f}%
\mathcal{S}$ can be elaborated for describing accelerating expansion (in
general, in locally anisotropic form) of the Universe, see details and
reviews of results in \cite{nojod1,eegrg,vbubuianu17}.

\subsection{Effective scalar field and STQC inflation}

The inflationary paradigm solved the problems of Big Bang cosmology by
adding a period of nearly exponential expansion during early time
cosmological evolution. In the literature there many historical and
comprehensive reviews for various MGTs and cosmology see \cite%
{capoz,nojod1,gorbunov,clifton,basilakos,linde2,gwyn,sami,mavromatos,nojiri,vagnozzi}
and references therein. In the simplest approach, such inflation scenarios
were modeled by homogeneous solutions of (modified) Einstein and nonlinear
Klein-Gordon equations for scalar fields.

In this subsection, we provide a condensed description and analyze some
concrete toy models of inflation with nontrivial STQC structure. For
simplicity, we assume that the inflationary dynamics is controlled by
certain nonlinear gravitational and matter field interactions resulting in
STQC configurations when the main issues can be addressed in some diagonal
parameterizations or approximations for a perfect (effective) matter fluid
encoding time and space quasiperiod structures and memory of off-diagonal
nonlinear interactions. For small parametric constructions with quasi FLRW
metrics (\ref{targdm4c})-(\ref{targdm4f}), our approach can be related to
cosmological perturbation theories but we do not discuss such details here.

\subsubsection{Scalar field description and gravitational STQC inflation}

Consider a scalar field with $\phi (x^{i},\tau )$ with action (\ref{actm})
for $~^{m}\mathcal{L=}\frac{1}{2}(\mathbf{e}_{\alpha }\phi )(\mathbf{e}%
^{\alpha }\phi )+V(\phi )$ and compute parameterize the energy-momentum
tensor $\ ^{m}\mathbf{T}_{\alpha \beta }$ (\ref{emtm}) for a diagonalized
metric (\ref{qflrewtqc}) with $\overline{\check{\eta}}_{i}(\tau )\simeq
1+\varepsilon \check{\chi}_{i}(\tau ),$ as follow
\begin{equation*}
\ ^{m}\mathbf{T}_{\acute{j}}^{\acute{\imath}}=p\delta _{\acute{j}}^{\acute{%
\imath}},\mbox{ for }p=\frac{1}{2}(\phi {^{\bullet }})^{2}-V(\phi ),\ ^{m}%
\mathbf{T}_{4}^{4}=-\rho =\frac{1}{2}(\phi {^{\bullet }})^{2}+V(\phi ).
\end{equation*}%
These formulas result in modified Friedmann equations $\frac{3}{2M_{P}^{2}}%
\check{H}^{2}=\rho $. Taking the first derivative on $\tau ,$ considering
previous formula for $\rho ,$ and $\check{H}{^{\bullet }=-}M_{P}^{2}(\phi {%
^{\bullet }})^{2},$ we obtain a nonlinear modification of the Klein-Gordon
equation for the canonical scalar filed $\phi $ in a nonholonomic
cosmological background,%
\begin{equation*}
\phi {^{\bullet \bullet }+3}\check{H}\phi {^{\bullet }+\partial }V/\partial
\phi =0.
\end{equation*}%
This allows us to compute gravitational STQC modifications of inflation of
the slow-roll indices $\ ^{\inf }\epsilon $ and $\ ^{\inf }\eta $ (we use
left abstract labels "inf" from inflation in order to avoid ambiguities with
other type values also denoted by $\epsilon $ and, respectively, $\eta ),$%
\begin{eqnarray*}
~^{\inf }\epsilon &=&-\frac{\check{H}{^{\bullet }}}{\check{H}^{2}}=~^{\inf }%
\mathring{\epsilon}\frac{1+(\ln \sqrt{|\overline{\eta }|}){^{\bullet \bullet
}/}\mathring{H}{^{\bullet }}}{[1+(\ln \sqrt{|\overline{\eta }|}){^{\bullet }/%
}\mathring{H}]^{2}}\mbox{ for }~^{\inf }\mathring{\epsilon}=-\frac{\mathring{%
H}{^{\bullet }}}{\mathring{H}^{2}}; \\
~^{\inf }\eta &=&-\frac{\check{H}{^{\bullet \bullet }}}{2\check{H}\check{H}{%
^{\bullet }}}=~^{\inf }\mathring{\eta}\frac{1+(\ln \sqrt{|\overline{\eta }|})%
{^{\bullet \bullet \bullet }/}\mathring{H}{^{\bullet \bullet }}}{2[1+(\ln
\sqrt{|\overline{\eta }|}){^{\bullet }/}\mathring{H})][1+(\ln \sqrt{|%
\overline{\eta }|}){^{\bullet \bullet }/}\mathring{H}{^{\bullet }}]},\ %
\mbox{for }~^{\inf }\mathring{\eta}=-\frac{\mathring{H}{^{\bullet \bullet }}%
}{2\mathring{H}\mathring{H}{^{\bullet }}}.
\end{eqnarray*}%
The slow-roll conditions for inflation state that $\ ^{\inf }\epsilon ,\
^{\inf }\mathring{\epsilon},~^{\inf }\eta ,~^{\inf }\mathring{\eta}\ll 1,$
when $\phi {^{\bullet \bullet }}\ll \check{H}$ $\phi {^{\bullet }}$ and $%
\phi {^{\bullet \bullet }}\ll \mathring{H}\phi {^{\bullet }}$, the modified
Friedman and Klein equations, and respective formulas for slow-roll indices,
are simplified as follows,
\begin{eqnarray*}
\check{H}^{2} &\simeq &\frac{2M_{P}^{2}}{3}V\mbox{ and }3\check{H}\phi
^{\bullet }\simeq -{\partial }V/\partial \phi , \\
\ ^{\inf }\epsilon &=&-M_{P}^{2}\frac{(\phi {^{\bullet })}^{2}}{\check{H}^{2}%
},\ ^{\inf }\mathring{\eta}=2M_{P}^{2}\frac{(\phi {^{\bullet })}^{2}}{\check{%
H}^{2}}+2\frac{\phi {^{\bullet \bullet }}}{\check{H}\phi {^{\bullet }}}.
\end{eqnarray*}

Using pervious formulas, we can compute the relations between prime and
target slow-roll indices and the potential $V(\phi )$ following formulas,%
\begin{eqnarray}
~^{\inf }\epsilon &=&~^{\inf }\mathring{\epsilon}\frac{1+(\ln \sqrt{|%
\overline{\eta }|}){^{\bullet \bullet }/}\mathring{H}{^{\bullet }}}{[1+(\ln
\sqrt{|\overline{\eta }|}){^{\bullet }/}\mathring{H}]^{2}}\simeq \frac{1}{%
4M_{P}^{2}}\left( \frac{{\partial }V/\partial \phi }{V}\right) ^{2},
\label{inflsrf} \\
~^{\inf }\eta &=&~^{\inf }\mathring{\eta}\frac{1+(\ln \sqrt{|\overline{\eta }%
|}){^{\bullet \bullet \bullet }/}\mathring{H}{^{\bullet \bullet }}}{2[1+(\ln
\sqrt{|\overline{\eta }|}){^{\bullet }/}\mathring{H})][1+(\ln \sqrt{|%
\overline{\eta }|}){^{\bullet \bullet }/}\mathring{H}{^{\bullet }}]}\simeq
\frac{1}{2M_{P}^{2}}\frac{|{\partial }^{2}V/\partial ^{2}\phi |}{V}.  \notag
\end{eqnarray}%
The modified graceful exit from the inflationary era occurs when $\ ^{\inf
}\epsilon \sim \mathit{O}(1).$ Using formulas (\ref{inflsrf}), we can
compute nonholonomic deformations by gravitational STQC structures of the
spectral indices from a prime state into target one, for
\begin{eqnarray*}
\mathring{n}_{s} &\simeq &1-6~^{\inf }\mathring{\epsilon}+2~^{\inf }%
\mathring{\eta},~^{\inf }\mathring{\eta}\simeq 16~^{\inf }\mathring{\epsilon}%
, \\
n_{s} &\simeq &1-6~^{\inf }\epsilon +2~^{\inf }\eta ,~^{\inf }\eta \simeq
16~^{\inf }\epsilon ,
\end{eqnarray*}%
calculated at the horizon crossing. We can consider various cosmological
configurations with nontrivial vacuum structure determined by a
gravitational polarization $\overline{\eta }[x^{i},y^{a};\overline{\varsigma
},\overline{\chi },\overline{\breve{q}},\overline{\psi },\overline{q},...]$.
For corresponding functional dependencies, such configurations may induce an
inflationary era, or to result in a graceful exit.

\subsubsection{Inflation determined by a scalar field with TC structure}

\label{ssinfls1tc}We can consider instead the Lagrangian for a scalar filed
the Lagrange density (\ref{1tqc}). In the non-relativistic approximation $%
\varsigma \rightarrow \varsigma (t),$ we can summarize the motion equations
and the Klein-Gordon equations for $\varsigma (t)$ and consider a nonlinear
equation,
\begin{equation*}
(\varsigma ^{\bullet })^{2}\varsigma {^{\bullet \bullet }+3}\check{H}%
\varsigma {^{\bullet }+2}\frac{\partial \acute{V}}{\partial \varsigma }=0.
\end{equation*}%
A subclass of solutions can be constructed for $\varsigma ^{\bullet }\simeq
1+\varepsilon \chi ^{\bullet }$ with a small parameter $\varepsilon $ and $%
\acute{V}\sim \varepsilon V(\chi )$, which results in
\begin{equation*}
\chi {^{\bullet \bullet }+3}\check{H}\chi {^{\bullet }+\partial Z(\chi
)/\partial \chi }=0,
\end{equation*}%
with redefined potential $\partial Z(\chi )/\partial \chi ={2}\frac{\partial
V(\chi )}{\partial \chi }+{3}\check{H}.$ In result, we obtain modifications
of the inflation slow-roll parameters (\ref{inflsrf}) by an additional TC
structure,%
\begin{eqnarray*}
~^{\inf }\epsilon &=&~^{\inf }\mathring{\epsilon}\frac{1+(\ln \sqrt{|%
\overline{\eta }|}){^{\bullet \bullet }/}\mathring{H}{^{\bullet }}}{[1+(\ln
\sqrt{|\overline{\eta }|}){^{\bullet }/}\mathring{H}]^{2}}\simeq \frac{1}{%
4M_{P}^{2}}\left( \frac{{\partial }Z/\partial {\chi }}{Z}\right) ^{2} \\
~^{\inf }\eta &=&~^{\inf }\mathring{\eta}\frac{1+(\ln \sqrt{|\overline{\eta }%
|}){^{\bullet \bullet \bullet }/}\mathring{H}{^{\bullet \bullet }}}{2[1+(\ln
\sqrt{|\overline{\eta }|}){^{\bullet }/}\mathring{H})][1+(\ln \sqrt{|%
\overline{\eta }|}){^{\bullet \bullet }/}\mathring{H}{^{\bullet }}]}\simeq
\frac{1}{2M_{P}^{2}}\frac{|{\partial }^{2}Z/\partial ^{2}{\chi }|}{Z}.
\end{eqnarray*}%
For such a model, the inflation era is determined both by an effective
one-dimensional time crystal structure for effective scalar fields and
nonlinear gravitational fields.

\subsubsection{STQC inflation from $F(R)$ theories}

We analyse slow-roll inflation for an example of MGT with STQC structure
when the evolution dynamics is determined by four generalized slow-roll
indices. Such values can be introduced and computed for off-diagonal
solutions as in \cite{nojod1,eegrg,vbubuianu17} using $~^{i}\check{H}$ (\ref%
{hablcdef}),%
\begin{eqnarray*}
\check{\epsilon}_{1} &=&-\frac{\check{H}{^{\bullet }}}{\check{H}^{2}}=%
\mathring{\epsilon}_{1}\frac{1+(\ln \sqrt{|\overline{\eta }|}){^{\bullet
\bullet }/}\mathring{H}{^{\bullet }}}{[1+(\ln \sqrt{|\overline{\eta }|}){%
^{\bullet }/}\mathring{H}]^{2}},\mbox{ for }\mathring{\epsilon}_{1}=-\frac{%
\mathring{H}{^{\bullet }}}{\mathring{H}^{2}}; \\
\check{\epsilon}_{2} &=&0;\check{\epsilon}_{3}\simeq -\check{\epsilon}_{1}%
\mbox{ and }\mathring{\epsilon}_{3}\simeq -\mathring{\epsilon}_{1};\check{%
\epsilon}_{4}\simeq -3\check{\epsilon}_{1}+\check{H}^{-1}(\ln |\check{%
\epsilon}_{1}|){^{\bullet }\mbox{ and }\mathring{\epsilon}_{4}\simeq -3%
\mathring{\epsilon}_{1}+\mathring{H}^{-1}(\ln |\mathring{\epsilon}_{1}|){%
^{\bullet }}.}
\end{eqnarray*}%
Such values allow us to compute in the slow-roll limit (when $\check{\epsilon%
}_{1},\check{\epsilon}_{4}\ll 1$) the nonholonomic $\overline{\eta }$%
--deformations to a target gravitational STQC configuration when respective
spectral index of primordial curvature perturbations and the
scalar-to-tensor ratio can be computed following formulas
\begin{subequations}
\begin{equation}
\check{n}_{s}\simeq 1-6\check{\epsilon}_{1}-2\check{\epsilon}_{4}{%
\mbox{ and
}\check{r}=48(\check{\epsilon}_{1})}^{2}.  \label{obsindmg}
\end{equation}%
We can generate a viable inflation model for a target configuration even
with observational compatible $\check{n}_{s}$ and ${\check{r}}$ even certain
primary data $\mathring{\epsilon}_{1}$ and ${\mathring{\epsilon}_{4}}$ do
not satisfy necessary conditions for inflation era with similar $\mathring{n}%
_{s}$ and ${\mathring{r}}$.

The Starobinsky $R^{2}$ model of inflation \cite{starob} was studied in a
number of cosmological works because it is compatible with Planck data \cite%
{plank,plank1} and can explain equivalently a scalar field and a modified
gravity model involving a phenomenological constant $~^{i}H$ of dimension of
$[mass^{2}].$ In hour approach \cite%
{cosmv2,eegrg,cosmv3,rajpvcosm,amaral17,vbubuianu17}, we consider that
during an effective slow-roll inflation era
\end{subequations}
\begin{equation}
\mathbf{F}(\widehat{\mathbf{R}})=\widehat{\mathbf{R}}+\widehat{\mathbf{R}}%
^{2}/36(~^{i}H),  \label{quadrmodif}
\end{equation}%
where constants a chosen in such a form which allow for LC-configurations to
apply the method for solution of FLRW equations following formulas
(306)-(316) in \cite{nojod1} and generalized to N-adapted computations in
\cite{eegrg,vbubuianu17}. For a flat quasi FLRW metric (\ref{qlestqc}) with
small $h$-polarizations, the Friedman equations are equivalent to
\begin{eqnarray}
\check{H}{^{\bullet \bullet }-}\frac{(\check{H}{^{\bullet })}^{2}}{2\check{H}%
}+3\check{H}(~^{i}H+\check{H}{^{\bullet })} &=&0,  \label{friedeq} \\
\widehat{\mathbf{R}}{^{\bullet \bullet }+3}\widehat{\mathbf{R}}(\check{H}%
+2~^{i}H) &=&0.  \notag
\end{eqnarray}%
During a slow-roll era, we can neglect the first two terms in the first
equation in the system above and find a quasi de Sitter solution for
cosmological quasi linear in time $\tau $ evolution when
\begin{equation*}
\check{H}(\tau )\simeq ~^{0}\check{H}-~^{i}H(\tau -\tau _{k}),
\end{equation*}%
where the STQC nonholonomic deformations are encoded in
\begin{equation}
\ ^{0}\check{H}=~^{0}\mathring{H}[1+(\ln \sqrt{|~^{0}\overline{\eta }|}){\
^{\bullet }/}~^{0}\mathring{H}].  \label{encod1}
\end{equation}%
Here we consider a prime a Hubble constant $~^{0}\mathring{H}$ modified by $%
\ ^{0}\overline{\eta }{^{\bullet }}(\tau _{k}),$ when $\tau $ takes values
from a time instance $\tau _{k},$ when small-roll approximations hold true,
till a moment $\tau =\tau _{f}$ for $\check{\epsilon}_{1}(\tau _{f})\simeq
1, $ when
\begin{equation*}
\check{H}(\tau _{f})=\check{H}_{f}\simeq \sqrt{~^{i}H}{\mbox{ and }}\tau
_{f}-\tau _{k}\simeq ~^{0}\check{H}/~^{i}H.
\end{equation*}%
Here we note that both $~^{0}\check{H}$ and $~^{i}H$ are expected to have
large values during inflation era (at least for target models). The
effective $e$-folding number is introduced by definition and computed
\begin{equation}
\check{N}:=\int_{\tau _{k}}^{\tau _{f}}\check{H}(\tau )d\tau \simeq (~^{0}%
\check{H})^{2}/2~^{i}H.  \label{efold}
\end{equation}%
This number is affected by the STCQ structure via $\ ^{0}\overline{\eta }$,
see formula (\ref{encod1}). Usually, there are considered large $\check{N}$
limits in the observational formulas.

Using above formulas, we can find STQC deformations of the observational
indices (\ref{obsindmg}) when%
\begin{eqnarray}
\check{n}_{s} &{\simeq }&1-4~^{i}H(~^{0}\check{H}-\frac{2~^{i}H\check{N}}{%
~^{0}\check{H}})^{-2}\simeq 1-\frac{2}{\check{N}},\mbox{ for large }\check{N}%
,  \notag \\
&\simeq &1-\frac{2}{~^{0}\check{N}[1+(\ln \sqrt{|~^{0}\overline{\eta }|}){\
^{\bullet }/}~^{0}\mathring{H}]^{2}}\mbox{ for }~^{0}\check{N}\simeq (~^{0}%
\mathring{H})^{2}/2~^{i}H;  \label{nindinf} \\
{\check{r}} &\simeq &{48(~^{i}H)}^{2}(~^{0}\check{H}-\frac{2~^{i}H\check{N}}{%
~^{0}\check{H}})^{-4}\simeq \frac{12}{\check{N}^{2}},\mbox{ for large }%
\check{N},\simeq \frac{12}{(~^{0}\check{N})^{2}[1+(\ln \sqrt{|~^{0}\overline{%
\eta }|}){^{\bullet }/}~^{0}\mathring{H}]^{4}}.  \label{rindinf}
\end{eqnarray}%
In Einstein and Jordan N-adapted frames, we yield the same observation
indices at the leading order for target metrics and models. The graceful
exit from inflation in such MGTs occurs due to the $R^{2}$ term, it
distortion, and STQC polarizations. All such effects may contribute to
un-stability of the de Sitter point. The vacuum de Sitter configurations are
with memory and STQC structure when nonlinear perturbations could result in
unstable locally anisotropic quasi de Sitter attractors. Hence, the graceful
exit comes as a result of growing curvature perturbations, nonholonomic
deformations (which can make stabile the dynamics for certain special
classes of non-integrable constraints) and because of gravitational and
matter filed quasi-periodic interactions and pattern-forming structures.

\subsubsection{Reheating and STQC structures in nonholonomic MGTs}

In this section, we discuss in brief how STQC structures affect the
reheating period in $R^{2}$ gravity (\ref{quadrmodif}) (the most
representative modified gravity model) that fills the gap between inflation
and the radiation and matter domination era. In our approach, ${~^{i}H}$ is
the same as $H_{i}$ in formulas (482)-(498) from review \cite{nojod1}. Those
results can be used to study observational indices of inflation, see
formulas (\ref{nindinf}) and (\ref{rindinf}); the reheating Friedman
temperatures. During the reheating era, the term $\widehat{\mathbf{R}}{%
^{\bullet \bullet }}$ in (\ref{friedeq}) became important and the evolution
of the scalar curvature is driven by an equation which is similar for that
for damped harmonic oscillator with restoring force of order $3{~^{i}H}$.

Reheating is a conventional oscillatory era when the terms $\check{H}{%
^{\bullet \bullet }}$ and $\frac{(\check{H}{\ ^{\bullet })}^{2}}{2\check{H}}$
dominate but $3\check{H}\check{H}{^{\bullet }}$ sub-dominates in the first
equation (\ref{friedeq}). Here we note that during the reheating phase $%
\widehat{\mathbf{R}}{\simeq 6}\check{H}{^{\bullet }}$ which is in contrast
with $\widehat{\mathbf{R}}{\simeq 6}\check{H}^{2}$ for the slow-roll era.
This results in such solutions for the reheating phases:%
\begin{eqnarray*}
\check{a} &=&\check{a}_{r}[1+\frac{\omega }{4}(\tau -\check{\tau}%
_{r})]^{2/3},\ \check{H}{\simeq }\frac{8\omega }{3}\frac{\cos ^{2}\omega
(\tau -\check{\tau}_{r})}{8+2\omega (\tau -\check{\tau}_{r})+\sin 2\omega
(\tau -\check{\tau}_{r})}, \\
\widehat{\mathbf{R}} &{\simeq }&\frac{16\omega }{3}\frac{\sin 2\omega (\tau -%
\check{\tau}_{r})}{8+2\omega (\tau -\check{\tau}_{r})+\sin 2\omega (\tau -%
\check{\tau}_{r})},
\end{eqnarray*}%
where $\check{a}_{r}=a_{0}\exp [(~^{0}\check{H})^{2}/2~^{i}H-1/12]$ for a
scale factor $a_{0}$ corresponding to the onset of inflation, $\omega =\sqrt{%
3~^{i}H/2}$ and the reheating time $\check{\tau}_{r}{\simeq (}~^{0}\check{H}%
)(~^{i}H).$ The values $\check{a}_{r}$ and $\check{\tau}$ are affected by a
STQC structure via $~^{0}\check{H}$ (\ref{encod1}) with a memory $~^{0}%
\overline{\eta }$ but the parameter $~^{i}H$ and angular velocity for
oscillations are not subjected to direct modifications.

Let us explain how the Universe is reheated both by the nonholonomic and
STQC structure and oscillations during such a phase. We can use the formulas
(490)-(492) in \cite{nojod1} rewritten for "inverse hat" values and averaged
square of a respective scalar curvature $\overline{\mathbf{\hat{R}}},$ in
our case, for the canonical d-connection $\widehat{\mathbf{D}}$. This
redefine the system (\ref{friedeq}) for the $(\tau ,\tau )$ component in the
form%
\begin{eqnarray*}
\check{H}^{2}+\frac{\widehat{\mathbf{R}}\check{H}}{18~^{i}H}(~1-\frac{%
\widehat{\mathbf{R}}^{2}\check{H}}{12}{)} &{=}&\frac{8\pi }{3}G(\ ^{m}\rho ),
\\
\widehat{\mathbf{R}}{^{\bullet \bullet }+3}\widehat{\mathbf{R}}(\check{H}%
+2~^{i}H) &=&\nu \frac{G(~^{i}H)\omega}{8\check{H}}\overline{\mathbf{\hat{R}}%
}^{2}.
\end{eqnarray*}%
where $G$ is the gravitational constant, $\nu $ is the number of matter
fields involved in the reheating process (during the reheating era, we
exclude the massless conformal fields which are not excited) and the energy
density of matter is computed
\begin{equation*}
\ ^{m}\rho =\frac{\nu \omega }{1152\pi }\frac{1}{\check{a}^{4}}\int_{\check{%
\tau}_{r}}^{\tau }\overline{\mathbf{\hat{R}}}^{2}\check{a}^{4}d\tau {\simeq }%
\frac{3}{5}\frac{32}{1152\pi }\frac{\nu \omega ^{3}}{\tau -\check{\tau}_{r}};
\end{equation*}%
for cosmic times $\tau \gg \check{\tau}_{r}+\omega ^{-1},$ the matter energy
$\ ^{m}\rho $ tends to zero and we obtain radiation dominating solutions.
The effective $\ ^{m}\rho $ is slightly affected by STQC structure via $%
\check{\tau}_{r}$ but this dependence is lost for cosmic times. In result,
we can estimate the corresponding reheating and Friedmann temperatures as in
$R^{2}$ gravity with standard FLRW metrics,
\begin{equation*}
T_{r}{\simeq 24\times 10}^{17}\sqrt{~^{i}H}l_{Pl}Gev\mbox{ and }T_{F}\sim
\nu ^{-3/4}{10}^{17}(36~^{i}Hl_{Pl}^{2})^{3/4}Gev,
\end{equation*}%
see formulas (494)-(498) in \cite{nojod1} for speculations on constraints on
parameter $~^{i}H$ following form the conditions that the reheating
temperature has to be large enough in order to allow the baryogenis, remain
monopole free etc.

The main conclusion of this subsection is that possible off-diagonal and
STQC configurations modelled with quasi FLRW metrics do not modify
substantially the reheating era in MGTs and GR. We can consider hidden quasi
periodic gravitational dark energy and dark matter structures which
contribute substantially to galactic formation etc. but such configurations
do not affect substantially reheating processes.

Finally, we note that using effective data $\check{H},\check{a},\widehat{%
\mathbf{R}}$ encoding off-diagonal and/or nonholonomic STQC configurations,
we can elaborate on various models of inflation, for instance, in mimetic $%
F(R)$ theories of gravity, topological Gauss-Bonne and other type
modifications of the GR, see review of such theories in Part III of \cite%
{nojod1}. MGTs result in very different cosmological scenarios of inflation
with STQC structure. We omit such considerations in this work.

\section{Late STQC dynamics, dark energy and dark matter}

\label{s5}The $\Lambda$CDM (LCDM) model is important in modern cosmology
because it fits a series of observational data and predicted the location
and existence of the baryon acoustic oscillations. In this section, we study
this model for (modified) gravity with STQC structure and discuss some
phenomenological aspects of dark energy era (i.e. late-time evolution) of
the Universe.

\subsection{Quasi $\Lambda $CDM epoch from STQC structures in $F(R)$ gravity}

For MGTs, this issue is addressed and reviewed in \cite{nos,degos,nojod1}.
It is also studied in the context of off-diagonal cosmological solutions and
quasiperiodic structures in \cite%
{cosmv2,eegrg,cosmv3,rajpvcosm,amaral17,vbubuianu17}. We show how that
approach with reconstruction techniques can be generalized for STQC
configurations with effective data $\check{H}(\tau ),\check{a}(\tau ),%
\widehat{\mathbf{R}}(\tau ),$ when the computations are performed for a flat
quasi FLRW metric (\ref{qflrewtqc}) when $\overline{\check{\eta}}_{i}(\tau
)\simeq 1+\varepsilon \check{\chi}_{i}(\tau )$ as in (\ref{targdm4f}). The
effective $\Lambda $CDM Hubble rate is $\check{H}^{2}-H_{0}^{2}=\frac{2}{3}%
M_{P}^{2}\rho _{0}\check{a}^{-3},$ where $\check{H}$ and $\check{a}$ encode
possible STQC structure and $\rho _{0}$ and $H_{0}$ are constants, when $%
\check{\Lambda}=12H_{0}^{2}$ can be identified with $\check{\Lambda}=\
^{m}\Lambda +\ \overline{\Lambda }$ for nonlinear symmetries (\ref{nsym1b}).
We can consider $\rho _{0}$ as an integration constant for all (effective)
matter.

Prescribing any value of $\widehat{\mathbf{R}}[\overline{\eta }]$ encoding
various possible nonlinear contributions via gravitational polarization $%
\overline{\eta }$ (\ref{polarizf}), we can apply the reconstruction
technique \cite{nos}, see also section IV-B in \cite{nojod1} and \cite{eegrg}%
, we can encode STQC contributions into an effective $F(R)$ theory by
introducing a functional
\begin{equation*}
G(\check{N})=\check{H}^{2}(\check{N})=H_{0}^{2}+\frac{2}{3}M_{P}^{2}\rho
_{0}a_{0}^{-3}e^{-3\check{N}},
\end{equation*}%
where $a_{0}$ is also an integration constant and $\check{N}$ is an
e-folding function of type (\ref{efold}) expressed as a function of $%
\widehat{\mathbf{R}},$ $\check{N}=-\frac{a_{0}^{3}}{6M_{p}^{2}\rho _{0}}(%
\widehat{\mathbf{R}}-12H_{0}^{2})$. This formula can be used for elaborating
an analogous model with nonholonomic nonminimal coupling as in \cite{eegrg}
when the modified Einstein equations (\ref{nmceq}) are written
\begin{equation*}
6(\widehat{\mathbf{R}}-9H_{0}^{2})(\widehat{\mathbf{R}}-12H_{0}^{2})F^{%
\prime \prime }(\widehat{\mathbf{R}})-(\widehat{\mathbf{R}}%
-18H_{0}^{2})F^{\prime }(\widehat{\mathbf{R}})-F(\widehat{\mathbf{R}})=0.
\end{equation*}%
This equation can be solved analytically in terms of the Gauss
hypergeometric function $F(\widehat{\mathbf{R}})=F(\widehat{\alpha },%
\widehat{\beta },\widehat{\gamma },\widehat{r})$ when the parameters are
subjected to the conditions $\widehat{\alpha }+\widehat{\beta }=\widehat{%
\alpha }\widehat{\beta }=\widehat{\gamma }=-1/6$ and $\widehat{r}=-3+
\widehat{\mathbf{R}}/3H_{0}^{2}$.

Finally, we note that prescribing STQC configurations we can reproduce
various models of effective $F(\widehat{\mathbf{R}})$ gravity explaining a
series of observational data for $F(\widehat{\mathbf{R}})\sim R^{2}$.
Following Starobinsky work \cite{starob}, such theories can be modelled
equivalently via interactions with a scalar field. In this work, we show
that certain classes of solutions can be parameterized to describe
quasiperiodic configurations.

\subsection{STQC unification of inflation with dark energy era in MGTs}

In this section we speculate on unified description of the inflationary era
with the dark energy, DE, era in the context of $F(R)$ gravity. For
LC-configurations, such results are reviewed in \cite{nojod1}. In this
sections, we study certain conditions for elaborating cosmological theories
with nontrivial STQC structure. In the late-time era with effective STQC
structure, we take
\begin{equation}
\check{F}(\widehat{\mathbf{R}})=\widehat{\mathbf{R}}-2\check{\Lambda}(1-e^{-%
\widehat{\mathbf{R}}/R_{0}})+\ ^{i}\Lambda (1-e^{-(\widehat{\mathbf{R}}/\
^{i}R)^{\varsigma }})+\check{\gamma}\widehat{\mathbf{R}}^{\check{\alpha}}.
\label{expmg}
\end{equation}%
In this formula, $\check{\Lambda}=\ ^{m}\Lambda +\ \overline{\Lambda }$ and
certain constants $R_{0},\ ^{i}\Lambda ,\ ^{i}R,\check{\gamma},\check{\alpha}
$ will be determined from the conditions that STQC configurations are
modelled equivalently both in GR and a suitable MGT and resulting in
observational data. This way, the quasiperiodic structure can preserved in
the limit $\widehat{\mathbf{R}}\gg R_{0}$ and realized by nonholonomic
deformations or LC-configurations of the Einstein-Hilbert action and
satisfying stability of the late-time de Sitter vacuum and avoiding
antigravity. For such an approximation, the local corrections to the Newton
law are negligible.

The essential features of the STQC $F(R)$-model can be studied using the
(effective) DE evolution $\check{\rho}_{DE}=\check{\rho}_{eff}-\rho /\check{F%
}^{\prime }(\widehat{\mathbf{R}})$, where $\rho $ is for the standard matter
fields and the quasi FLRW equation is written
\begin{equation}
\check{\rho}_{eff}=\frac{3}{2M_{P}^{2}}\check{H}^{2}=[\rho +(2M_{P})^{-2}(%
\check{F}^{\prime }\widehat{\mathbf{R}}-\check{F}-6\check{H}(\check{F}%
^{\prime }){^{\bullet }})]/\check{F}^{\prime }.  \label{qfeq2}
\end{equation}%
The conservation laws for DE are considered in a standard effective fluid
form,%
\begin{equation*}
\frac{d\check{\rho}_{DE}}{d(\ln \check{a})}+3(\check{\rho}_{DE}+\check{p}%
_{DE}),
\end{equation*}%
for a EoS parameter $\check{w}_{DE}=\check{p}_{DE}/\check{\rho}_{DE}.$

To analyze properties of the STQC F-modified Friedmann equation (\ref{nmceq}%
) is convenient to introduce a new variable and redefine the parameters,
\begin{equation*}
\check{y}:=\check{\rho}_{DE}/\rho _{m}^{(0)}=\check{H}^{2}/\tilde{m}^{2}-%
\check{a}^{-3}-\chi \check{a}^{-4},
\end{equation*}%
where the mass scale $\tilde{m}^{2}=2M_{p}^{2}\rho _{m}^{(0)}/3\simeq
1.5\times 10^{-67}eV^{2}$ is computed for the energy density of matter at
present time, $\rho _{m}^{(0)},$ and $\chi =\rho _{r}^{(0)}/\rho
_{m}^{(0)}\simeq 3.1\times 10^{-4}$ is determined by the energy density of
radiation at present time. Approximating $\widehat{\mathbf{R}}=12\check{H}%
^{2}+6\check{H}{^{\bullet }}$ for $\check{H}=\ln |\check{a}|$, we express (%
\ref{qfeq2}) for a functional $\check{y}(\check{H})$ in the form%
\begin{equation*}
\frac{d^{2}\check{y}}{(d\check{H})^{2}}+\check{J}_{1}\frac{d\check{y}}{d%
\check{H}}+ \check{J}_{2}\check{y}+\check{J}_{3}=0.
\end{equation*}%
In this equation, there are considered functions
\begin{eqnarray*}
\check{J}_{1} &=&4+\frac{1-\check{F}^{\prime }(\widehat{\mathbf{R}})}{6%
\tilde{m}^{2}\check{F}^{\prime \prime }(\widehat{\mathbf{R}})(\check{y}+%
\check{a}^{-3}+\chi \check{a}^{-4})},\check{J}_{2}=\frac{2-\check{F}^{\prime
}(\widehat{\mathbf{R}})}{3\tilde{m}^{2}\check{F}^{\prime \prime }(\widehat{%
\mathbf{R}})(\check{y}+\check{a}^{-3}+\chi \check{a}^{-4})}, \\
\check{J}_{3} &=&-3\check{a}^{-3}-\frac{1-\check{F}^{\prime }(\widehat{%
\mathbf{R}})}{6\tilde{m}^{2}\check{F}^{\prime \prime }(\widehat{\mathbf{R}})}%
+\frac{\check{F}(\widehat{\mathbf{R}})-\widehat{\mathbf{R}}}{18\tilde{m}^{4}%
\check{F}^{\prime \prime }(\widehat{\mathbf{R}})(\check{y}+\check{a}%
^{-3}+\chi \check{a}^{-4})},
\end{eqnarray*}%
when the scalar curvature and effective EoS parameter are determined by STQC
structure encoded in $\check{y}$ as follows,
\begin{equation*}
\widehat{\mathbf{R}}=3\tilde{m}^{2}(\frac{d\check{y}}{d\check{H}}+4\check{y}+%
\check{a}^{-3})\text{\mbox{ and }}\check{w}_{DE}=-1-\frac{1}{3}\frac{d\ln |%
\check{y}|}{d\check{H}}.
\end{equation*}

The constructions in this section are very similar to those in sections IV.A
and IV.C of \cite{nojod1}. In our approach, the difference is that our
variables $\check{y},\check{a},\check{H}$ etc. are stated by a STQC
structure with effective $\check{\Lambda}=\ ^{m}\Lambda +\ \overline{\Lambda
}$. It is also convenient to work with a different system of notations in (%
\ref{expmg}). We can use those numeric solutions for a range $R_{0}\ll
\widehat{\mathbf{R}}\ll \ ^{i}R,$ with $R_{0}/\check{\Lambda}=0.6,0.8$ and
1, including the matter domination era and the current acceleration epoch
\begin{equation*}
\check{\Lambda}=\ ^{m}\Lambda +\ \overline{\Lambda }=7.93\tilde{m}^{2},\ \
^{i}R/2=\ \ ^{i}\Lambda =10^{100}\check{\Lambda},\varsigma =4,\check{\alpha}%
=5/2,\check{\gamma}=(4\ \ ^{i}\Lambda )^{1-\check{\alpha}}.
\end{equation*}%
Here we note that such results can be reproduced by choosing the initial
conditions in a form similar to that for the standard $\Lambda $CDM model
for a red shift $\check{z}=\check{a}^{-1}-1.$ There are taken such initial
conditions for $\check{z}=\ ^{i}\check{z}$ when $(d\check{y}/d\check{z})_{|\
^{i}\check{z}}=0$ and $\check{y}_{|\ ^{i}\check{z}}=(3\tilde{m}^{2})^{-1}%
\check{\Lambda}.$ There is a compatibility with WMAP results if the values $%
\ ^{i}\check{z}=1.5,2.2,2.5$ are chosen respectively for $R_{0}/\check{%
\Lambda}=0.6,0.8$ and 1 and $\widehat{\mathbf{R}}=3\tilde{m}^{2}(\check{z}%
+1)^{3}.$ The respective value of the EoS parameter $\check{z}=0$ (present
epoch) result respectively in values: $-\check{w}_{DE}=0.994,0.975,0.950,$
which for the DE is very closed to the phantom line $-1.$ Such results prove
that even the STQC structure may substantially change the vacuum and
nonvacuum cosmological evolution via nontrivial gravitational polarization $%
\overline{\eta }$ (\ref{polarizf}) we can chose such effective variables and
parameters when the models are equivalent to certain $F(R)$ theories or
solutions in describing GR with off-diagonal nonlinear interactions.

\section{Conclusions}

\label{s6}With this article we aimed to elaborate on a new class of
inflation and late time cosmological models with time quasiperiodic
structures in MGTs and GR. We tried to make the paper self-consistent, so we
reviewed in brief our geometric method on constructing generic off-diagonal
locally anisotropic cosmological solutions and analyzed possible
nonholonomic constraints and parametric decompositions resulting in diagonal
and effective $\Lambda $CDM models encoding space-time quasiperiodic, STQC,
structure. Following results of Planck2015 \cite{plank,plank1} (that the
ration of tensor perturbation amplitude $r<0.1$), a number of researchers
concluded that such observational data seem to "virtually eliminate all the
simplest textbook inflationary models". In order to solve this problem and
up-date cosmological scenarios, theorists elaborate on MGTs and constructing
new classes of exact and parametric solutions. Former methods of
transforming generalized Einstein equations into systems of nonlinear
ordinary differential equations with solutions determined by integration
constants fail in elaborating accelerating cosmology scenarios with
quasiperiodic structure and locally anisotropic and inhomogeneous evolution.

It is well-known that modern cosmological data (see details and discussions
in \cite{linde2,nojod1,capoz,nojiri,cosmv2,nos,degos,plank,plank1}) prove that
the acceleration of the Universe is determined by a new dark energy, DE, and
dark matter, DE, physics. This resulting in existence of  complex network \
filament, quasiperiodic and aperiodic like structures (with various
nonlinear wave interactions, diffusion processes,  fractal like
configurations etc.). Such "hidden" and directly observed rich geometric
gravitational field/ spacetime and (effective) matte structures are driven
by  a nonlinear cosmological evolution  with respective scale
dependence for meta-galactic and galactic configurations. It is considered
that such modified cosmological models can be elaborated realistically only
for a new class of MGTs even for certain cases the GR seem to be correct if,
for instance,  nonlinear locally anisotropic and inhomogeneous cosmological solutions with generic
off-diagonal terms for metrics are considered. The formation and evolution
of cosmological configurations with nonlinear gravitational and matter field
structures can be modelled by numeric, analytic and geometric methods as
deformations of an effective quasycristal, QC, space structure for the
gravitational and fundamental scalar fields, see  \cite%
{cosmv3,amaral17,vbubuianu17} and references therein. It should be
emphasized that we can not describe quasiperiodic and pattern forming
structures as exact solutions of (modified) Einstein equations if we work
 with cosmological models based only on the Friedman-Lama\^{\i}%
tre-Robertson-Worker, FLRW, metric and/or standard Bianchi anisotropic
generalizations with metrics depending only a time like variable and
strcutre Lie group symmetries. Using such "simplified" ansatz for metrics,
we can transform the gravitational and matter field equations into certain
systems of nonlinear ordinary differential equations, ODEs, and certain
algebraic matrix equations. Such equations can be solved in terms of
integration constants but this does not allow, for instance, to obtain
nonlinear/ solitonic waves (or other type cosmological complex structures
like QC, networks  and filaments) which involve generic off-diagonal
solutions on nonlinear partial differential equations, PDEs, expressed in
terms of generation and integration functions, parametric nonlinear
dependencies etc. We have to elaborate more advanced geometric methods in order to
solve analytically such physically important systems of PDEs.  Using only numeric and approximate methods result in  "non-exact" solutions which do not provide a complete understanding of properties of corresponding
nonlinear dynamics.

Our anholonomic frame deformation method, AFDM, \cite%
{cosmv2,eegrg,cosmv3,vbubuianu17} allows to construct more realistic classes
of exact and parametric solutions determined by generating and integration functions and
(effective) matter sources containing, in general, possible modifications of
GR from classical and quantum gravity theories. The advantage of the AFDM \
is that it allows us to chose generating functions and sources in various
forms describing   STQC-, QC-, pattern forming, nonlinear wave etc.
structures as exact and parametric cosmological solutions of generalized
Einstein equations both in MGTs and GR. In general, such solutions are
inhomogeneous and locally anisotropic and described by generic off-diagonal
metrics, non-Riemann connections, with coefficients depending, in principle,
on all spacetime coordinates and subjected to certain nonholonomic
constraints and generalized symmetries. Nevertheless, it is always possible
to impose at the end some additional constraints (for instance,
the condition of zero torsion, certain observable boundary
behaviour and symmetries, or to consider homogeneous and isotropic
configuration with nontrivial topology) and search for possible limits to  known cosmological scenarios. We  emphasize here that if we impose certain fixed high symmetry conditions and chose a "very simplified" ansatz from the very beginning (before finding generalized solutions for nonlinear systems of
PDEs), the possibility to construct general self-consistent cosmological
scenarios with quasiperiodic and pattern forming structures is lost. Such
configurations can not be completely described in terms of ODEs and their
particular solutions. We have to involve more general classes of solutions
of (modified) Einstein equations transformed into corresponding effective
PDEs in order to explain and predict modern cosmology observational and
experimental data.

The novelty of this work is that we developed the AFDM in a form which
allows to generate cosmological STQC-type structures both for the
gravitational and matter fields (which is important for elaborating new
theories of DE and DM physics) as exact and parametric solutions. Our new
classes of cosmological solutions prove that nonlinear gravitational and
matter field interactions induce quasiperiodic structures which are
generalizations (in certain approximations, similar) of time crystals
constructed as analogous to space crystals \cite{wilczek12} but those models
where elaborated for generalized nonlinear mechanics, classical and quantum
dynamical systems and condensed matter physic and not for gravity theories
and cosmology.  For quasiperiodic space configurations, such structures model
respective spacetime QC structures as we elaborated in our previous works
\cite{cosmv3,amaral17,vbubuianu17}. The primary emphasis of quantum theory
in formatting classical time crystals in solid state physics is analyzed in
\cite{shapere12}. In our work (it is the first one on STQC), we study how analogous
configurations can generated as (off-) diagonal cosmological configurations
in  MGTs and  GR. In particular, it is shown that various (quasi) periodic
gravitational and matter field structures can be generated by a spontaneous
broken gravitational vacuum with different nonlinear symmetries.  We can
motivate that the early universe can be described by a STQC gravitational
dynamics and filled, for instance, with DM time quaisperiodic structures by
considering a more complete nonlinear cosmological dynamics determined by
exact solutions of generalized Einstein equations. Such exact solutions can
be redefined in arbitrary systems of reference/coordinates. So, prescribing
gravitational/matter field generating functions with time-dependence and
quasiperiodicity (for instance, in a system of reference where the system of
field equations were decoupled in a general form)  we can recompute such
values for any choice of a time coordinate in a physical system of reference.
This is possible because the AFDM method allows to integrate physically
important systems of PDEs in general exact or parametric forms and to
consider such solutions in any geometric/ physical system of reference. If
we would work with approximate numeric methods, the type of time dependence
and quasiperiodicity conditions would mix the coefficients of metrics and connections and affect
substantially the possibility of decoupling and integrating explicitly such nonlinear equations.
This is another important priority of the AFDM, see
\cite{cosmv2,eegrg,cosmv3,rajpvcosm,amaral17,vbubuianu17} and references
therein. Even there were published many papers on cosmological impacts of $%
f(R)$- and $f(T)$-theories, the dynamics of the universe etc. described by other
classes of less general solutions, those constructions could not prove the existence and encode
STQC-structures.  In sections  \ref{s4} and \ref{s5}, we elaborated
respectively on such time quasiperiodic inflationary and acceleration
cosmology models and proved that such STQC-models describe more realistically modern observational
data.

In condensed matter physics, the subject of dynamical fluctuations of time
crystals was investigated in \cite{castillo13,castillo14}. We refer to those
works for detailed explanations on how effective field theories can be
derived. In our case, we study the self-consistent small parametric
evolution of quantum fluctuations which can be encoded as locally
anisotropic cosmological structures modelling space-time crystals. Such
dynamical systems posses a local conventional lowest energy and performs a
periodic motion. STQC cosmological dynamics involve a more rich nonlinear
pattern forming structure, with memory of nonlinear effects. This can be
used for more realistic elaboration of dark energy and dark matter models.

Finally, we note that our approach is based on the hypothesis that inflation
is primarily driven by vacuum energy at a scale indicated by gauge coupling
unification \cite{hertzberg17}. Concretely, in this paper, we elaborated a
class of models with hybrid inflations for which the vacuum energy is
associated with a grand unified theory condensate. Such a quasi-periodic
analogous condensate provides the dominant energy during inflation and a
second inflation scalar slow-rolls. In a more general context, we can
consider quantum corrections and modified dispersion relations which is a
goal for our future research.

\vskip5pt \textbf{Acknowledgments:}\ This work consists a natural extension
of the research program for the project IDEI in Romania,
PN-II-ID-PCE-2011-3-0256. Author thanks D. Singleton for hosting his adjunct
position at Fresno State University. He is also grateful for references on
time crystals (related to A. Shapere and F. Wilczek works \cite%
{shapere12,wilczek12}) and important discussions on former works on locally
anisotropic quasiperiodic structures in (modified) gravity and cosmology
\cite{cosmv2,eegrg,cosmv3,vbubuianu17}.

\end{document}